\documentclass[preprints,article,accept,oneauthor,pdftex]{myCls} 
\firstpage{1} 
\pubvolume{1}
\issuenum{1}
\articlenumber{0}
\pubyear{2022}
\copyrightyear{2022}
\datereceived{} 
\dateaccepted{} 
\datepublished{} 
\hreflink{https://doi.org/} 
\newcommand{\ib}[1]{{\color{black}#1}}

\newcommand{\alt}{\raisebox{-0.3ex}{$\stackrel{<}{\sim}$}} 

\newcommand{\figname}{Figure~}

\newcommand{\Figsname}{Figures~}
\newcommand{\figsname}{Figures~}
\newcommand{\secname}{Section~}
\newcommand{\secsname}{Sections~}

\newcommand{\Gl}{Equation}
\newcommand{\Gls}{Equations}
\usepackage[version=4]{mhchem} 
\usepackage{threeparttable}
\usepackage{extarrows}
\DeclareGraphicsExtensions{.jpg, .eps, .pdf, .png}
\graphicspath{{/media/ioan/Elements/gt/1graphene/.}{.}}
\Title{Estimating the Number of Molecules in Molecular Junctions Merely Based on the Low Bias Tunneling Conductance at Variable Temperature}
\TitleCitation{Estimating the Number of Molecules in Molecular Junctions Merely Based on the Low Bias Tunneling Conductance at Variable Temperature}
\Author{{Ioan B\^aldea} 
}
\AuthorNames{Ioan B\^aldea}

\AuthorCitation{B\^aldea, I.}
\address[1]{%
Theoretical Chemistry, Heidelberg University, Im Neuenheimer Feld 229, D-69120 Heidelberg, Germany; ioan.baldea@pci.uni-heidelberg.de 
}

\abstract{
  Temperature ($T$) dependent conductance $G = G(T)$ data measured in molecular junctions are routinely taken as evidence for a two-step hopping
  mechanism. The present paper emphasizes that this is not necessarily the case.
  A curve of $\ln G$ versus $1/T$ decreasing almost linearly (Arrhenius-like regime) and eventually switching to a nearly horizontal plateau
  (Sommerfeld regime), or possessing a slope gradually decreasing with increasing $1/T$ is fully compatible with a single-step tunneling mechanism.
  The results for the dependence of $G$ on $T$ presented include
  both analytical exact and accurate approximate formulas and numerical simulations.
  \ib{These theoretical results are general, also in the sense that they are not limited, e.g., to the 
    (single molecule electromigrated (SET) or large area EGaIn) fabrication platforms,
    which are chosen for exemplification merely in view of the available experimental data needed for analysis.}
  To be specific, we examine in detail transport measurements for molecular junctions based on ferrocene (Fc).
  As a particularly important finding, we show how the present analytic formulas for $G=G(T)$ can be utilized to
  compute the ratio $f = A_{\text{eff}} / A_n$ between the effective and nominal areas of large area Fc-based junctions with
  an EGaIn top electrode. Our estimate of $f\approx 0.6 \times 10^{-4}$ is comparable with previously reported values \ib{based on completely different methods}
  for related large area molecular junctions.
}
\begin{document}
\section{Introduction}
\label{sec:intro}
Comparing charge transport properties of single molecule junctions
with junctions based on ensembles of molecules represents an
important issue that has been frequently addressed in the past \cite{Frisbie:03,McCreery:09,McCreery:13a,Guo:16b,Frisbie:16d,Cahen:20}.
The former includes \ib{mechanically controllable break junctions \mbox{(MCBJ) \cite{Reed:97,Loertscher:07}}}
and scanning tunneling microscopy (STM) break junctions
\cite{Reichert:02,Tao:03,Venkataraman:06,Ruitenbeek:08}
as well as electromigration \cite{Reed:09,Reed:11,Nijhuis:16b} platforms. 
\textls[-10]{Conducting probe atomic force microscopy (CP-AFM) \cite{Frisbie:00,Frisbie:01,Frisbie:02,Frisbie:02b,Frisbie:04},
\mbox{cross-wires \cite{Kushmerick:02,Kushmerick:02b,Kushmerick:05,Beebe:06,Beebe:08}}}
and large area liquid metal (eutectic gallium indium alloy EGaIn) based molecular \mbox{junctions \cite{Whitesides:13,Whitesides:14,Nijhuis:16b,Zhao:22}}
are examples of the second category.
For the latter, the key role played by the number of molecules and the related effective contact area
has been thoroughly emphasized in the literature \cite{Whitesides:13,Frisbie:16d,Park:19b}. To exemplify, let us refer
to the variation of the low bias conductance $G$---the property (\ib{determined experimentally from the slope of $I$-$V$ curve at low biases where the curve is linear)}
which will be in the main focus below---across a homologous molecular family whose members contain a variable number of repeat units $n$.
Claiming the ubiquitous exponential decay
$G_n = G_C \exp(-\tilde{\beta} n)$ \cite{Frisbie:00,Frisbie:01,Tao:03,Frisbie:04,Guo:11,Wandlowski:08c,Frisbie:11} 
by monitoring values of conductance $G_n$ measured for junctions with various repeat units $n$
makes sense only if they contain the same number of molecules.
In the same vein, we can mention the tiny even-odd effect reported
in the tunneling decay coefficient $\tilde{\beta}$ and/or contact conductance $G_C$
\cite{Whitesides:11,Ramin:11,Whitesides:14c,Nijhuis:15b,Nijhuis:15f,Nijhuis:17c,BenAmara:20}.
The opposite claims on the direction of this tiny effect 
(odd members more conductive \cite{Whitesides:14,Nijhuis:17c}
versus even members more conductive \cite{Whitesides:11})
may reflect the difficulty of controlling the effective (``electric'') number of molecules
in large area junctions \cite{Selzer:05,Milani:07,Boer:08,Whitesides:13,Nijhuis:16g,Frisbie:16d,Cahen:17a,Cahen:20}.

Comparison between temperature dependent transport properties of junctions based on a CP-AFM platform \cite{Frisbie:16d} or
single-molecule transistors (SET) \cite{Nijhuis:16b} and large area junctions fabricated with EGaIn technique
using the same or similar molecular species has been attempted in the past to address the 
issue of effective, ``electric'' area versus nominal, ``geometric'' area.
Nonetheless, the inherently different nature of the contacts of EGaIn- and, e.g., CP-AFM-based junctions
(EGaIn top electrode versus AFM metal coated tip) raises some difficulty in interpreting the results of this certainly meaningful approach.

As elaborated below, the approach presented in this paper allows this difficulty to be obviated. It is merely based on
low bias conductance data collected on large area junctions at variable temperature.
The exact formula for the temperature dependent conductance in the low bias limit deduced recently by us \cite{Baldea:2022c}
constitutes the theoretical framework of this methodology, which is considered in the next section.
\section{{Results and Discussion}
}
\label{sec:results}
\subsection{General Results}
\label{sec:general}
According to the general Keldysh formalism \cite{Caroli:71a,Meir:92,HaugJauho,CuevasScheer:17},
the low bias conductance \linebreak \mbox{$G \equiv  \left . \partial I(V)\partial V\right\vert_{V\to 0} $}
of a single molecule tunneling junction at finite temperature \linebreak $\beta = \left(k_B T\right)^{-1}$ 
can be expressed as \cite{Baldea:2017d,Baldea:2022c}
\begin{equation}
  \frac{G}{G_0} = - \Gamma_{g}^2 \int_{-\infty}^{\infty}
  \frac{d\,\varepsilon}{\left(\varepsilon - \varepsilon_0\right)^2 + \Gamma_{a}^2} \frac{\partial}{\partial \varepsilon} f(\varepsilon)
  = \frac{\beta}{4}  \Gamma_{g}^2  \int_{-\infty}^{\infty} \frac{ \mbox{sech}^{2} (\beta \varepsilon / 2) }{\left(\varepsilon - \varepsilon_0\right)^2 + \Gamma_{a}^2}
  d\,\varepsilon
\label{eq-g}
\end{equation}
{Here }$G_0 \equiv 2 e^2/h = 77.48\,\upmu$S and $f(\varepsilon) = 1/\left(1 + e^{\beta \varepsilon}\right)$
are the quantum conductance and Fermi distribution, respectively, and energies are measured relative to
electrodes' Fermi energy ($E_F \equiv 0$).
In the present model, the charge transport is mediated by a single level (molecular orbital, MO),
and the coupling to two infinite wide, flat band $s$ (substrate) and $t$ (top, tip) electrodes
is quantified by an energy independent effective MO-electrode coupling $\Gamma_{g}$ \cite{Baldea:2021b},
\begin{equation}
  \Gamma_{g} \equiv \sqrt{\Gamma_s \Gamma_t }
  \label{eq-Gamma-g}
\end{equation}
which is the geometric average of the individual MO-electrode couplings $\Gamma_{s,t}$.
Effects due to charge image \cite{Sommerfeld:33,desjonqueres:96,Neaton:06,Baldea:2014a,Baldea:2014e}, gate potential \cite{Reed:09}, etc
that are responsible for level energy shifts are embodied in the renormalized value of $\varepsilon_0$, which is a model
parameter. In contrast to the isolated molecule, the embedded molecule has an MO 
possessing a finite energy width
\begin{equation}
  \Gamma_{a} \equiv \frac{1}{2}\left(\Gamma_s + \Gamma_t\right)
  \label{eq-Gamma-a}
\end{equation}

\Gl~(\ref{eq-g}) clearly emphasizes the two distinct impacts of $\Gamma_{s,t}$
on the tunneling transport. On one hand, they contribute multiplicatively via $\Gamma_g$ (cf.~\Gl~(\ref{eq-Gamma-g}))
as MO-electrode couplings that determine the overall magnitude of the tunneling current.
On the other hand, they contribute additively via $\Gamma_a$ (\Gl~(\ref{eq-Gamma-a}))
to the MO energy broadening, which can compete with the smearing of the electrodes' Fermi distributions
at \mbox{nonvanishing~temperatures}.

As shown recently \cite{Baldea:2022c}, the RHS of \Gl~(\ref{eq-g}) can be integrated out analytically.
The result for the conductance per molecule expressed via the real part of Euler's trigamma function of complex argument
function $\psi^{\prime}(z)$ \cite{AbramowitzStegun:64} reads
\begin{equation}
  \label{eq-g-exact}
\frac{G}{G_0} = 
\frac{\Gamma_{g}^2}{2 \pi \Gamma_{a} k_B T} \mbox{Re}\, \psi^{\prime} \left(\frac{1}{2} + \frac{\Gamma_{a}}{2\pi k_B T} + i\,\frac{\varepsilon_0}{2\pi k_B T}\right) 
\end{equation}
{The trigamma function} represents the derivative of the digamma function,
$\psi^{\prime}(z) \equiv \psi(1; z) \equiv \frac{d}{d\,z}\psi(z)$,
which in turn is the logarithmic derivative of Euler's gamma function
\cite{AbramowitzStegun:64}.
\mbox{\Gl~(\ref{eq-g-exact})} is an exact result valid at arbitrary values of all parameters
($\varepsilon_0$, $\Gamma_{g}$, $\Gamma_{a}$, and $T$).

\begin{figure}[H]
  {
    \includegraphics[width=0.5\textwidth,angle=0]{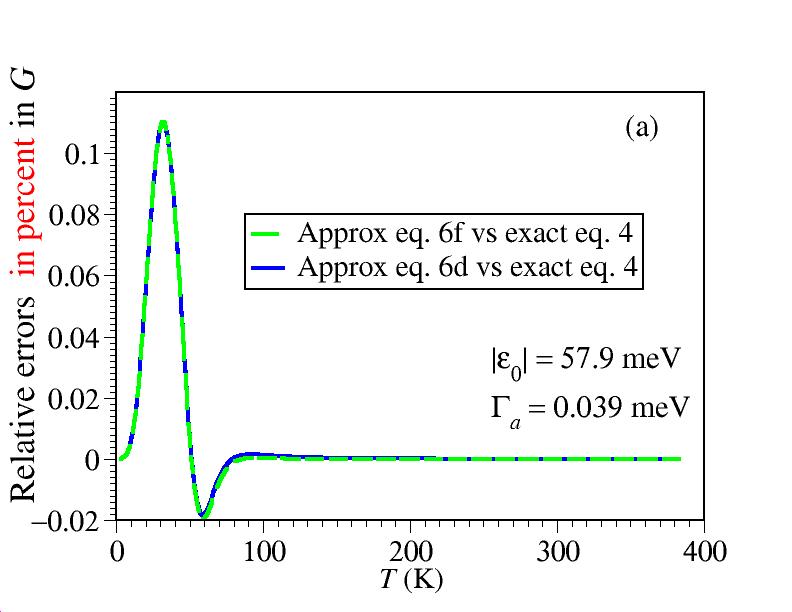}
    \includegraphics[width=0.5\textwidth,angle=0]{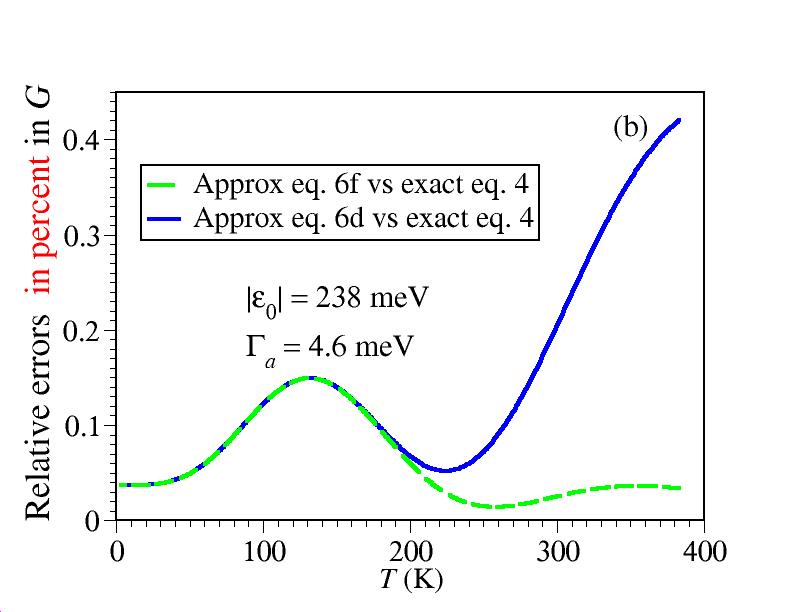}}
    \vspace{-12pt} 
  \caption{Relative errors in conductance $G$ for parameters (indicated in the legend) characterizing (\textbf{a}) the SET and (\textbf{b}) SAM setups (see below)
    illustrating that \Gl~(\ref{eq-g-approx-1},f)  represent very accurate approximation of the exact \Gl~(\ref{eq-g-exact}).}
  \label{fig:err1T-err2T}
\end{figure}

Noteworthily, $G$ does not depend on the sign of $\varepsilon_0$. 
The RHS of \Gl~(\ref{eq-g}) is invariant upon changing $\varepsilon_0 \to -\varepsilon_0$. This can 
easily be seen by changing the variable ($\varepsilon \to - \varepsilon$). 
Alternatively, this is also the consequence of the invariance of \Gl~(\ref{eq-g-exact}) under complex conjugation. 
Rephrased physically, junctions wherein conduction is mediated by LUMO ($\varepsilon_l = \varepsilon_0 > 0$)
and junctions wherein conduction is mediated by HOMO
($\varepsilon_h = - \varepsilon_0 < 0$) have the same conductance $G$. 
Using the analytic expression $\mbox{Im}\,\psi(1/2 + i y) = (\pi/2)\,\tanh (\pi y)$ \cite{AbramowitzStegun:64},
the lowest order Taylor expansion of the RHS of \Gl~(\ref{eq-g-exact}) yields
\begin{eqnarray}
  \label{eq-expansion}
  \mbox{Re}\,\psi^{\prime}\left(\frac{1}{2} + x + i y\right) & = &
  \frac{\pi^2}{2}\mbox{sech}^2\,(\pi y) + x\, \mbox{Re}\,\psi\left(2; \frac{1}{2} + i y\right) \nonumber \\
  & + & x^2 \frac{\pi^4}{2} \left[2 - \cosh(2 \pi y)\right] \mbox{sech}^4 (\pi y) + \mathcal{O}(x^3)
\end{eqnarray}
{The real part of the} tetragamma function $\psi(2; z) \equiv \frac{d^2}{d z^2} \psi(z)$ 
with $z=1/2 + i y$ (real $y$) entering above in the RHS is not available in closed analytic form;
however, we found that it
can be very accurately approximated (\figname\ref{fig:phi}) via elementary functions {as follows} 
 \cite{Baldea:2022c}
\begin{subequations}
  \label{eq-ReTetraGamma}
  \begin{eqnarray}
  \displaystyle
  & & \mbox{Re}\,\psi\left(2; \frac{1}{2} + i y\right) \simeq \varphi(y) \label{eq-g-approx} \\
  & & \varphi(y) = \frac{y^2 - 34.7298}{\left(y^2 + 2.64796\right)^2} +
  37.262 \frac{y^2 + 1.12874}{\left(y^2 +  2.17786\right)^3} + 3.01373 \frac{y^2 -0.082815}{\left(y^2 + 0.25014\right)^3} \label{eq-phi} 
\end{eqnarray}
\vspace{-12pt} 
\begin{figure}[H]
  {
    \includegraphics[width=0.5\textwidth,angle=0]{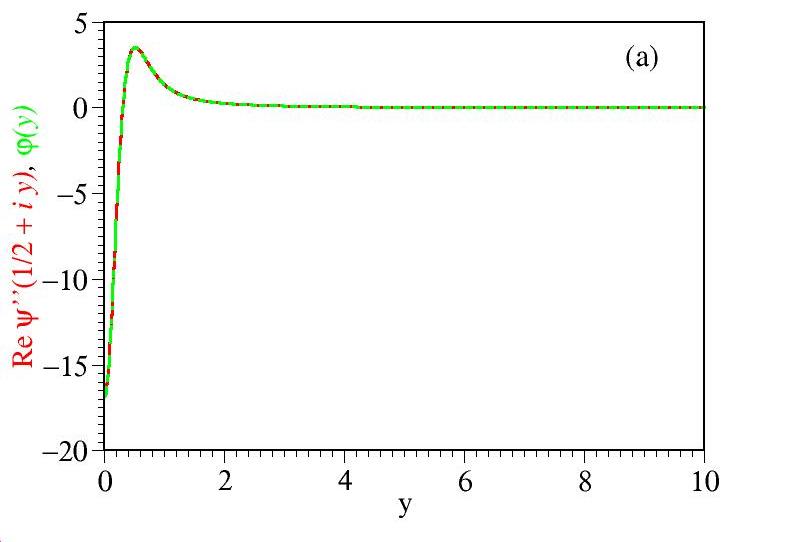}
    \includegraphics[width=0.5\textwidth,angle=0]{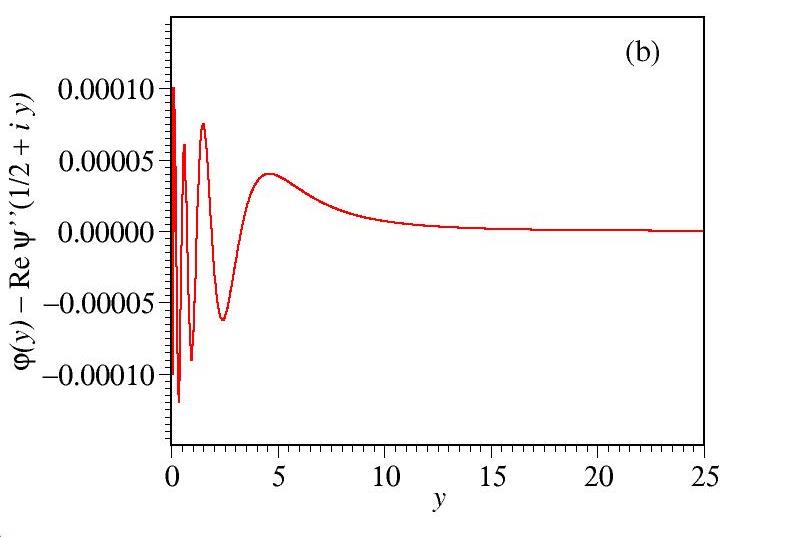}}
  \caption{(\textbf{a}) The function $\varphi(y)$ expressed in terms of elementary functions (\Gl~(\ref{eq-phi}))
    and the real part of the polygamma function $\mbox{Re}\,\psi(2; 1/2 + i y)$ depicted (\textbf{b}) along with their differences,
    revealing that \Gl~(\ref{eq-ReTetraGamma}) is a very accurate approximation.}
  \label{fig:phi}
\end{figure}
For parameter ranges covering virtually all experimental situations of interest wherein a $T$-dependent
$G$ can be expected, the parameter
\begin{equation}
  x \equiv \frac{\Gamma_{a}}{2\pi k_B T}
  \label{eq-x}
\end{equation}
is small, and the lowest order expansion of the RHS of \Gl~(\ref{eq-g-exact}) 
\begin{eqnarray}
  \displaystyle
  \frac{G}{G_0}
  &
  \simeq
  &
  \frac{\pi \Gamma_{g}^{2}}{4 \Gamma_{a} k_B T} \mbox{sech}^2 \pi y
  + \frac{\Gamma_{g}^2 \varphi(y)}{\left(2 \pi k_B T\right)^2}
  \label{eq-g-approx-1} \\
  y & \equiv & \frac{\varepsilon_0}{2 \pi k_B T} \label{eq-y}   
\end{eqnarray}
is a very accurate approximation of the exact \Gl~(\ref{eq-g-exact}); it holds $\mathcal{O}\left(x^2\right)$,
which amounts to an relative error of $\sim$1\% for $\Gamma_a$ smaller than about $k_B T/2$.
Notice the numerical factor 4 in the denominator of the first term of \Gl~(\ref{eq-g-approx-1}),
which corrects the incorrect factor 16 (a typo) in \Gl~(6) of ref.~\cite{Baldea:2022c}.
If (highly unlikely in real junctions exhibiting $T$-dependent transport) $x$ is not very
small with respect to unity, the last term in the RHS of \Gl~(\ref{eq-expansion})
can also be included
\begin{eqnarray}
  \displaystyle
  \frac{G}{G_0}
  &
  \simeq
  &
  \frac{\pi \Gamma_{g}^{2}}{4 \Gamma_{a} k_B T} \mbox{sech}^2 \pi y + \frac{\Gamma_{g}^2 \varphi(y)}{\left(2 \pi k_B T\right)^2} \nonumber \\
  & & + \frac{\pi}{16}\frac{\Gamma_{g}^2 \Gamma_{a}}{\left(k_B T \right)^3} \left[2 - \cosh (2 \pi y)\right] \mbox{sech}^4(\pi y);
  \label{eq-g-approx-2} 
\end{eqnarray}
\end{subequations}
it holds $\mathcal{O}\left(x^3\right)$,
which amounts to an relative error of $\sim$1\% for $\Gamma_a$ smaller than about $1.4\,k_B T$.
At temperatures lower than the aforementioned ($1.4\,k_B T~\alt~\Gamma_a$), thermal effects are negligible and
the zero temperature limit (\Gl~(\ref{eq-g0K})) applies.
\figname\ref{fig:err1T-err2T} illustrates the accuracy of the approximate \Gl~(\ref{eq-g-approx-1},f) 
for parameter values characterizing the real molecular junctions considered in \secname\ref{sec:real}.
The curves computed via \Gl~(\ref{eq-g-approx-1},f) 
cannot be distinguished within the drawing accuracy from those 
obtained via the exact \Gl~(\ref{eq-g-exact}) in \secsname\ref{sec:simulations} and \ref{sec:real}.
Therefore, they will not be shown there.
\vspace{-12pt} 

Noteworthily, \Gl~(\ref{eq-g-approx-1},f) only contain elementary functions.
This is important for practical data fitting; special functions like trigamma entering \Gl~(\ref{eq-g-exact}) are usually not implemented
in common software packages used by experimentalists.

For parameter values where the peaks of the transmission function and the derivative of the
Fermi function
---possessing widths of the order $\Gamma_{a}$ and $k_B T$, and
located at $\varepsilon = \varepsilon_0$ and $\varepsilon = 0$, respectively---are sufficiently well separated in energy, the following approximate~formula 
\begin{equation}
\displaystyle
\frac{G}{G_0} \simeq
 \underbrace{\frac{\pi}{4} \frac{\Gamma_{g}^2}{\Gamma_{a} k_B T} \mbox{sech}^2 \frac{\varepsilon_{0}}{2 k_B T}}_{G_T/G_0}
+ \underbrace{\frac{\Gamma_{g}^2}{\varepsilon_0^2 + \Gamma_{a}^2} }_{G_{0K}/G_0}
\xlongrightarrow{\Gamma_{a} \ll \left\vert\varepsilon_0\right\vert} 
\frac{\pi}{4} \frac{\Gamma_{g}^2}{\Gamma_{a} k_B T} \mbox{sech}^2 \frac{\varepsilon_{0}}{2 k_B T}
+ \frac{\Gamma_{g}^2}{\varepsilon_0^2}
\label{eq-g-pccp} 
\end{equation}
generalizes a result deduced earlier \cite{Baldea:2017d} for $\Gamma_s = \Gamma_t = \Gamma_{a} = \Gamma_{g}$.
It holds within $\sim$1\% for $\Gamma_a$ smaller than about $ k_BT/10$.
\Gl~(\ref{eq-g-pccp}) reduces in turn to \Gls~(\ref{eq-g0K}) and (\ref{eq-high-T})
in the limit of very low and very high temperatures, 
generalizing results known from earlier studies \cite{CuevasScheer:17,Baldea:2012g,Lambert:11,Nijhuis:16b,Baldea:2018a}.
\begin{equation}
  \frac{G}{G_0} \xlongrightarrow{k_B T \ll \Gamma_{a}} \frac{G_{0K}}{G_0}
  = \frac{\Gamma_{g}^2}{\varepsilon_0^2 + \Gamma_{a}^2} \xlongrightarrow{\Gamma_{a} \ll \vert \varepsilon_0 \vert}
  \frac{\Gamma_{g}^2}{\varepsilon_0^2}
  \label{eq-g0K}
\end{equation}
\begin{subequations}
  \label{eq-high-T}
\begin{equation}
\displaystyle
\label{eq-g-sech} 
\frac{G}{G_0} \xlongrightarrow{\Gamma_{a} \ll \pi k_B T} \frac{G_{T}}{G_0} \equiv
\frac{\pi}{4} \frac{\Gamma_{g}^2}{\Gamma_{a} k_B T} \mbox{sech}^2 \frac{\varepsilon_{0}}{2 k_B T} 
\end{equation}
\begin{equation}
  \displaystyle
  \label{eq-g-exp} 
  \frac{G}{G_0} \xlongrightarrow{\Gamma_{a} \ll \pi k_B T \ll \left\vert \varepsilon_0\right\vert }
  \frac{G_{p.A}}{G_0} = \frac{\pi\Gamma_{g}^2}{\Gamma_{a} k_B T} 
  \exp\left(-\frac{\left\vert\varepsilon_0\right\vert}{k_B T}\right)
\end{equation}
\end{subequations}
{(The above subscript $p.A$} stands for pseudo-Arrhenius).

Notice that unlike \Gl~(\ref{eq-g-pccp}), $T$ enters the RHS of \Gl~(\ref{eq-g-approx-1},f) 
not only in the first term but also in the second term.
Therefore, departures of \Gl~(\ref{eq-g-pccp}) from \Gl~(\ref{eq-g-exact}) become substantial
when $\vert\varepsilon_0\vert $, $\Gamma_{a}$, and $k_B T$ have comparable values.
For this reason, for temperatures around $T_c$ (see \Gl~(\ref{eq-Tc}) below), \Gl~(\ref{eq-g-exact})
better quantifies the gradual transition between
an Arrhenius-type (high $T$) and a Sommerfeld (low $T$) regime~\cite{Baldea:2022c} than \Gl~(\ref{eq-g-pccp}).

Thermal corrections to \Gl~(\ref{eq-g0K}) can alternatively obtained
via Sommerfeld expansion of \Gl~(\ref{eq-g}) and expressed in terms of the
Riemann $\zeta$ function \cite{Sommerfeld:33,JahnkeEmde:45,AshcroftMermin}
\begin{subequations}
   \label{eq-Sommerfeld}
\begin{adjustwidth}{-\extralength}{0cm}
 \begin{equation*}
    \frac{G}{G_0}  =  
    \frac{1}{4} \sum_{n=0}^{\infty} 
    \frac{\left(k_B T\right)^{2 n}}{(2 n) !}
    \left . \frac{\partial^{2 n}}{\partial\varepsilon^{2 n}}  \mathcal{T}(\varepsilon)\right\vert_{\varepsilon = 0}
    \int_{-\infty}^{\infty} x^{2 n} \mbox{sech}^2 \frac{x}{2} d\,x
    = \frac{G_{0K}}{G_0} + \sum_{n=1}^{\infty} \left(k_B T\right)^{2 n} \left(2 - \frac{1}{2^{2(n - 1)}}\right) \zeta(2 n)
  \end{equation*}
\end{adjustwidth}
  which gives the first Sommerfeld correction (S1, $\mathcal{O}\left(T^2\right)$)
  \begin{equation}
     \label{eq-Sommerfeld-1}
    \frac{G}{G_0} \simeq \frac{\Gamma_{g}^2}{\varepsilon_0^2 + \Gamma_{a}^2} \left[1 +
      \left(\pi k_B T\right)^2 \frac{\varepsilon_0^2 - \Gamma_{a}^{2}/3}{\left(\varepsilon_0^2 + \Gamma_{a}^2\right)^2}\right]
  \end{equation}
  and the second Sommerfeld correction (S2, $\mathcal{O}\left(T^4\right)$)
  \begin{equation}
     \label{eq-Sommerfeld-2}
    \frac{G}{G_0} \simeq \frac{\Gamma_{g}^2}{\varepsilon_0^2 + \Gamma_{a}^2} \left[1 +
      \left(\pi k_B T\right)^2 \frac{\varepsilon_0^2 - \Gamma_{a}^{2}/3}{\left(\varepsilon_0^2 + \Gamma_{a}^2\right)^2} +
      \left(\pi k_B T\right)^4 \frac{7}{15}
      \frac{5 \varepsilon_0^4 - 10 \varepsilon_0^2  \Gamma_{a}^{2} +  \Gamma_{a}^{4}}{\left(\varepsilon_0^2 + \Gamma_{a}^2\right)^4}
      \right]
  \end{equation}    
\end{subequations}

{Interestingly, there is } no linear correction in $T$ to $G$ in the above formulas.

\ib{To end this general theoretical part, and in order to avoid confusion regarding the applicability to real molecular junctions,
  we want to emphasize that none of the above results
  quantifying thermal effects on the charge transport via tunneling 
  is limited to a specific experimental platform, be it
  SET, EGaIn (to be examined in \secsname\ref{sec:real} and \ref{sec:N}),
  CP-AFM (considered earlier \cite{Baldea:2018a}) or any other.

  What is important for the single level model underlying \Gl~(\ref{eq-g}) is that the charge transport is ``one-dimensional'', i.e.,
  proceeds along \emph{{individual} 
} molecules;
loosely speaking, that an electron (or hole) leaving the left electrode does not tunnel across the left half of a molecule A,
then jumps on a neighboring molecule B, and finishes the trip to the right electrode after tunneling across the right half of molecule B.

Importantly, the theoretical single level model utilized does not necessarily rule out an intermolecular (A-B) interaction. 
In an elementary transport process, an electron tunneling across molecule A can interact with the adjacent molecule B.
Provided that the charge transport does not induce electron \emph{exchange} between adjacent molecules A and B,
the effects of this potentially significant intermolecular interaction translate into an extra level shift
(i.e., renormalized $\varepsilon_0$) and an
extra level broadening
expressed as an additional term to the RHS of \Gl~(\ref{eq-Gamma-a})
\begin{equation*}
  \Gamma_s + \Gamma_t \to \Gamma_s + \Gamma_t + \Gamma_{env}
\end{equation*}
{Above, the subscript} ``env'' stands for environment. Because both $\varepsilon_0$ and $\Gamma_a$ are model parameters,
the implications for data fitting are not dramatic.

The fact that in \secname\ref{sec:N} we will be able to estimate the fraction $f$ of active molecules merely in terms of $\Gamma_s$ and $\Gamma_t$
(amounting to assume $\Gamma_{env} = 0$) demonstrates that, at least for the large area EGaIn-based junctions considered there,
intermolecular interaction effects do not have a dramatical impact on transport.}
  
\subsection{Results Illustrating the Temperature Impact on the Charge Transport by Tunneling}
\label{sec:simulations}
\begin{figure}[H]
\begin{adjustwidth}{-\extralength}{0cm}
\centering 
{
    \includegraphics[width=0.43\textwidth,angle=0]{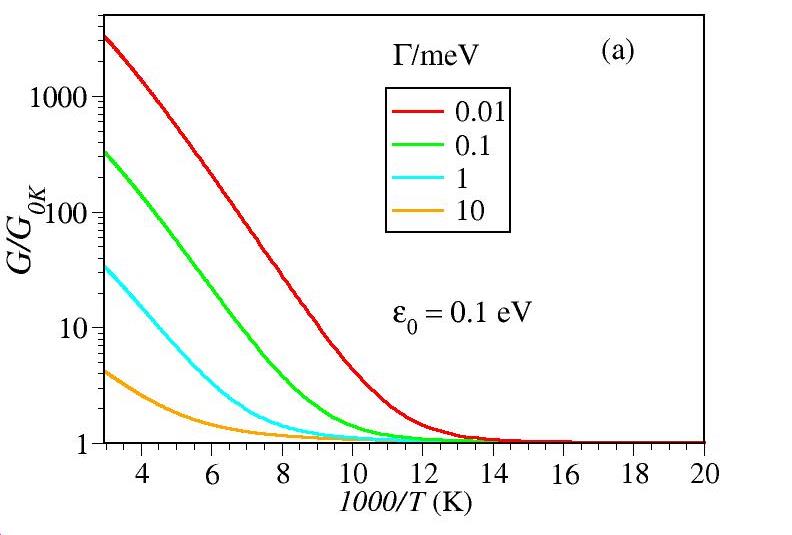}
    \includegraphics[width=0.43\textwidth,angle=0]{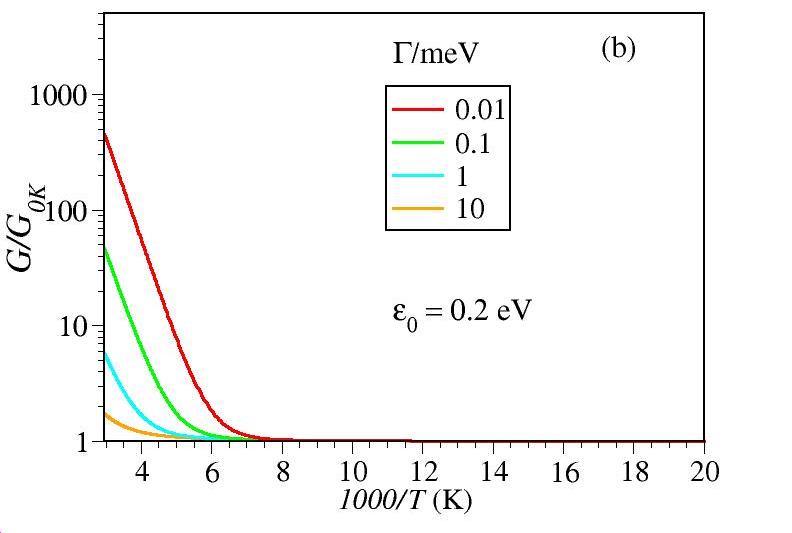}
    \includegraphics[width=0.43\textwidth,angle=0]{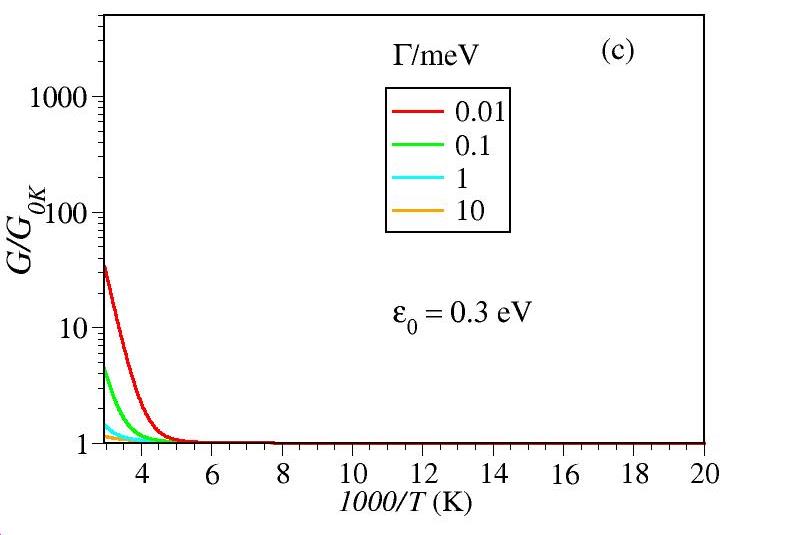}
  }
  \centerline{
    \includegraphics[width=0.43\textwidth,angle=0]{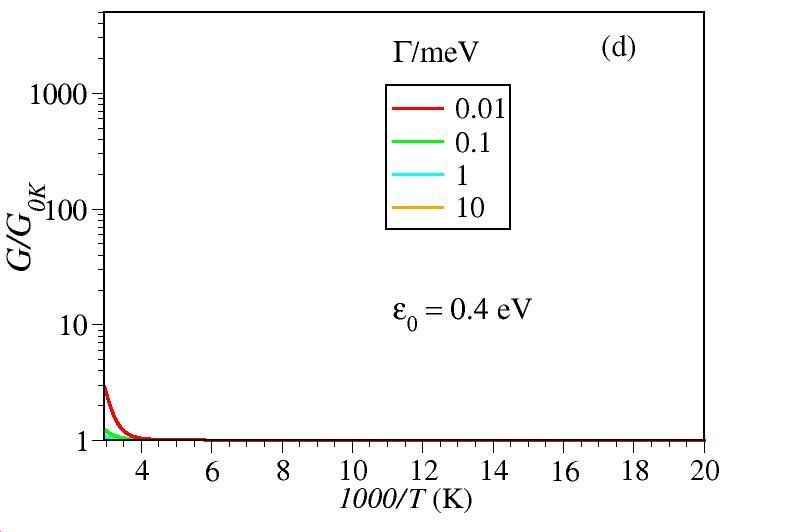}
    \includegraphics[width=0.43\textwidth,angle=0]{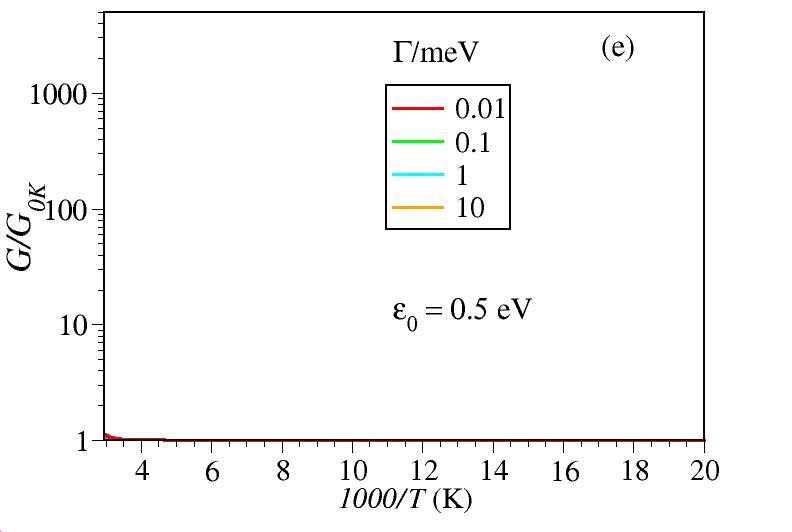}
    }
\end{adjustwidth}
  \caption{{Simulating the temperature }
impact on the low bias conductance $G$ normalized to the zero temperature
    value $G_{0K}$. Impact of a variable level width $\Gamma_{a}$ at fixed energy offset $\varepsilon_0$.
     Notice that the range on the \emph{y}-axis is the same all panels (\textbf{a}--\textbf{e}). At given $\varepsilon_0$, the impact of temperature
    is more pronounced at smaller $\Gamma_a$.
    As visible, thermal effects on $G$ are
  negligible for $\left\vert\varepsilon_0\right\vert$ larger than about 0.4\,eV.} 
  \label{fig:simul-e0}
\end{figure}

\begin{figure}[H]
  {
    \includegraphics[width=0.45\textwidth,angle=0]{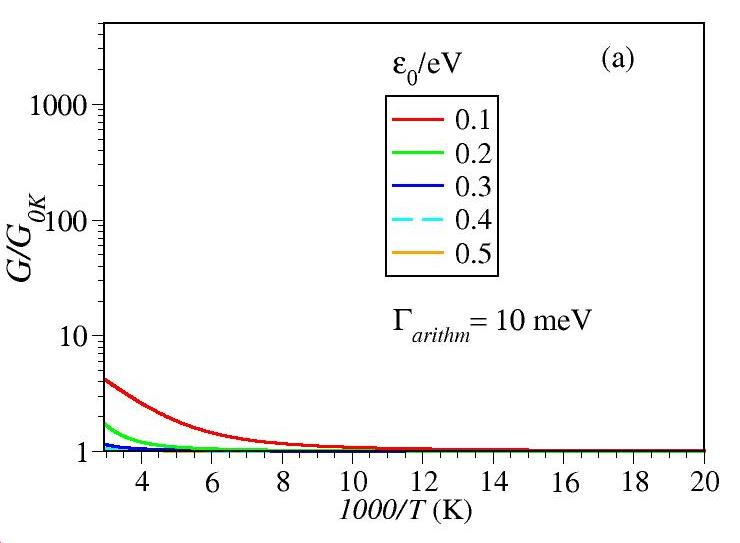}
    \unskip
    \includegraphics[width=0.45\textwidth,angle=0]{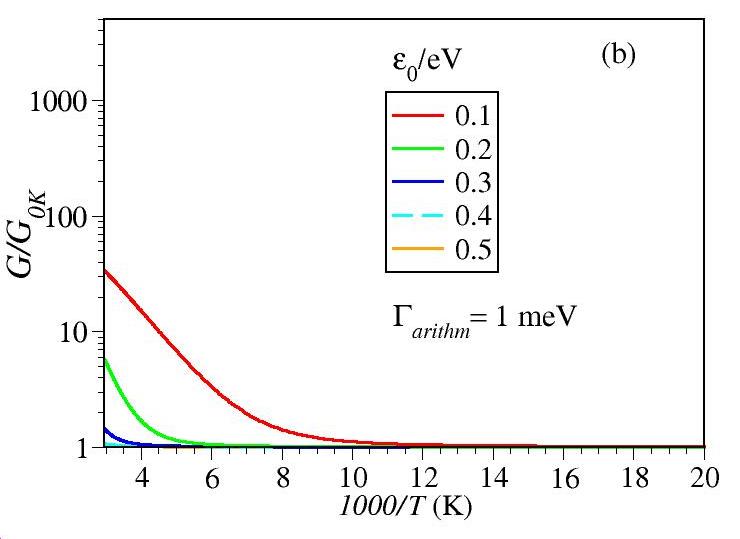}
  }
  {
    \includegraphics[width=0.45\textwidth,angle=0]{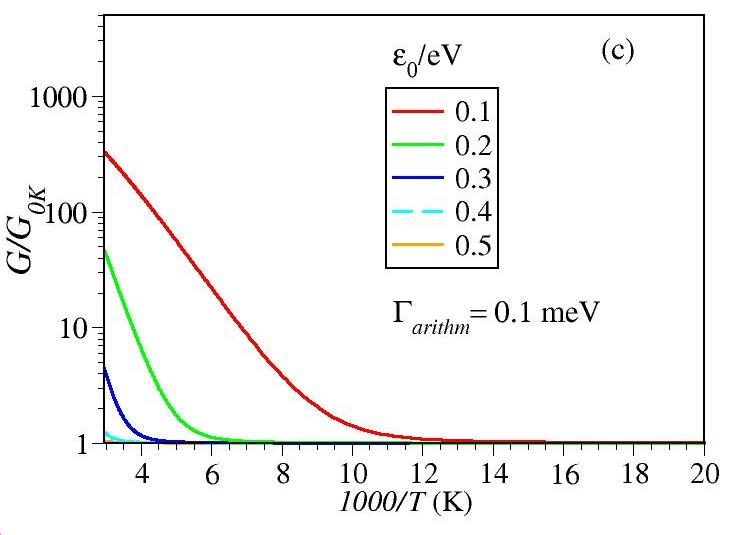}
    \includegraphics[width=0.45\textwidth,angle=0]{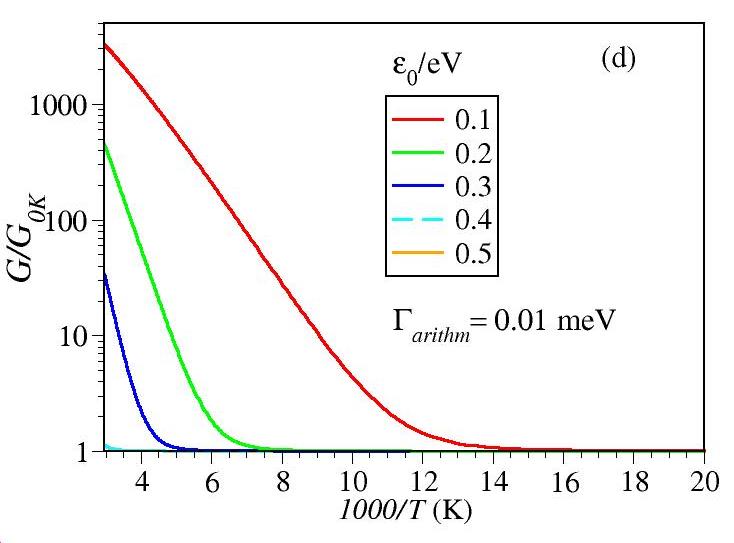}
    }
  \caption{Simulating the temperature impact on the low bias conductance $G$ normalized to the zero temperature
    value $G_{0K}$. Impact of a variable level offset $\varepsilon_0$ at fixed $\Gamma_{a}$.
     At given $\Gamma_a$, the impact of temperature
    is more pronounced at smaller $\left\vert\varepsilon_0\right\vert$.
    Notice that the range on the \emph{y}-axis is the same all panels (\textbf{a}--\textbf{d}).}
  \label{fig:simul-Delta}
\end{figure}

Insight into the thermal impact on the tunneling conductance can be gained by inspecting the results
of numerical simulations depicted in \figsname\ref{fig:simul-e0} and \ref{fig:simul-Delta}.
Inspection of these figures reveals that, irrespective of the magnitude of the MO width $\Gamma_{a}$, up 
to $T \approx 340$\,K---a value that safely covers the temperature range accessed in experiments \cite{Nijhuis:16b,Baldea:2018a}---thermal effects are negligible for energy offsets $\left \vert\varepsilon_0\right\vert$
larger than about $0.4$\,eV (cf.~\figname\ref{fig:simul-e0}d,e).

Below this value, thermal effects become significant. At a given level offset value $\left \vert\varepsilon_0\right\vert$,
they are the more pronounced, the smaller the value of $\Gamma_{a}$ is (cf.~\figname\ref{fig:simul-e0}a--c).
Likewise, at given level width $\Gamma_{a}$, thermal effects are the more pronounced, the smaller the level
offset  $\left \vert\varepsilon_0\right\vert$ is (cf.~\figname\ref{fig:simul-Delta}a--c).
%

By and large, the message conveyed by \figsname\ref{fig:simul-e0} and \ref{fig:simul-Delta} is clear: 
temperature dependent measured data
should by no means be taken as conclusive evidence for two-step hopping conduction
(cf.~ref.~\cite{Baldea:2022c} and citations therein).

\figsname\ref{fig:simul-e0} and \ref{fig:simul-Delta} clearly illustrate that, for sufficiently
(but realistically) small values of $\left \vert\varepsilon_0\right\vert$ and $\Gamma_{a}$
the single-step tunneling transport can exhibit a strong temperature dependence.
At high $T$, the (pseudo-)Arrhenius behavior resulting from tunneling (cf.~\Gl~(\ref{eq-g-exp})) 
can hardly be distinguished from the traditional Arrhenius characteristic for charge transport
via hopping. As the temperature is lowered, this Arrhenius-like regime $ G \approx G_T $
(cf.~\Gl~(\ref{eq-g-sech})) gradually switches into a
Sommerfeld regime \cite{Baldea:2022c}, wherein thermal effects basically represent corrections
(cf.~\Gl~(\ref{eq-Sommerfeld})) to the zero temperature value $G_{0K}$ (\Gl~(\ref{eq-g0K})).

\begin{figure}[H]
  {
    \includegraphics[width=0.45\textwidth,angle=0]{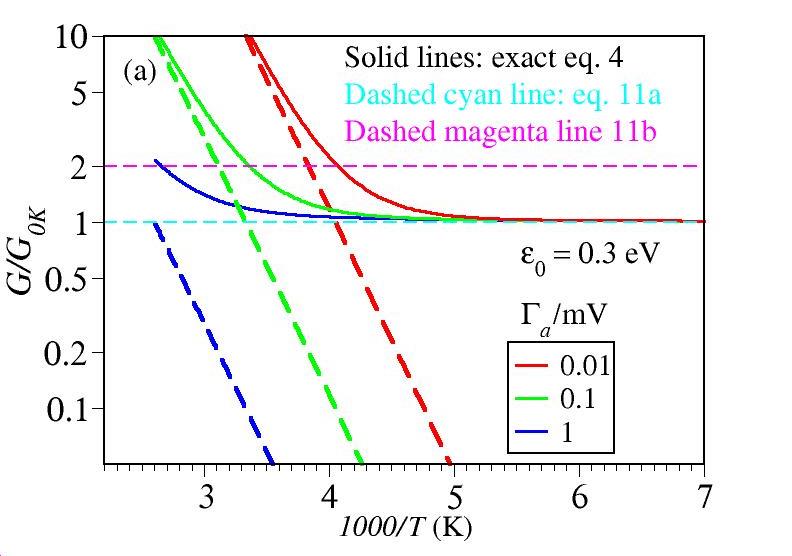}
    \unskip
    \includegraphics[width=0.45\textwidth,angle=0]{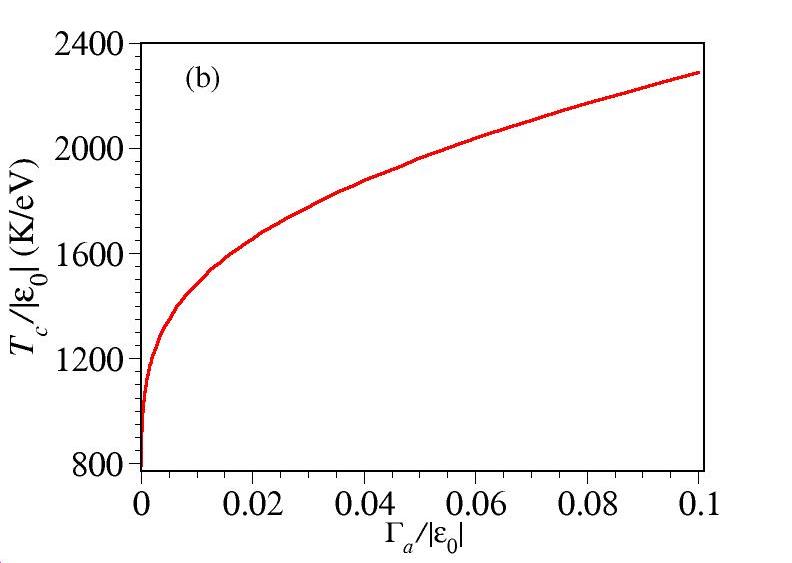}}
  {
    \includegraphics[width=0.45\textwidth,angle=0]{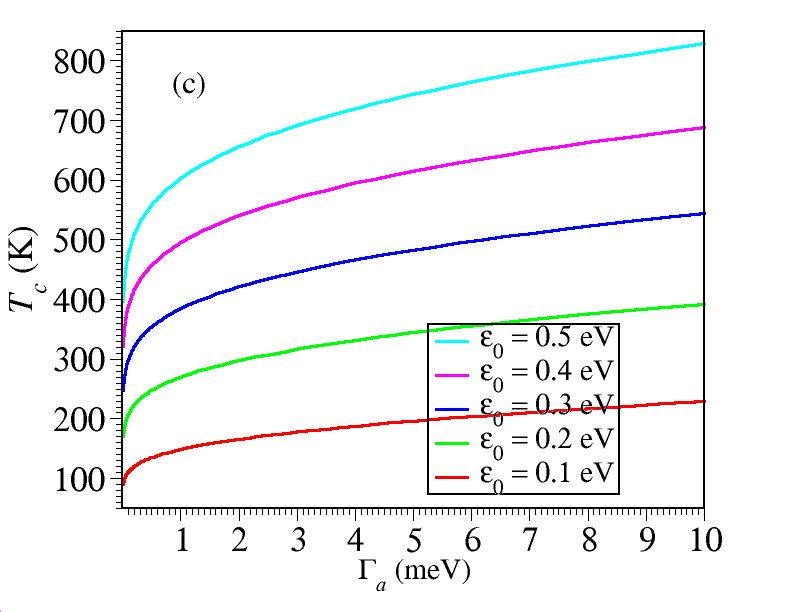}
    \includegraphics[width=0.45\textwidth,angle=0]{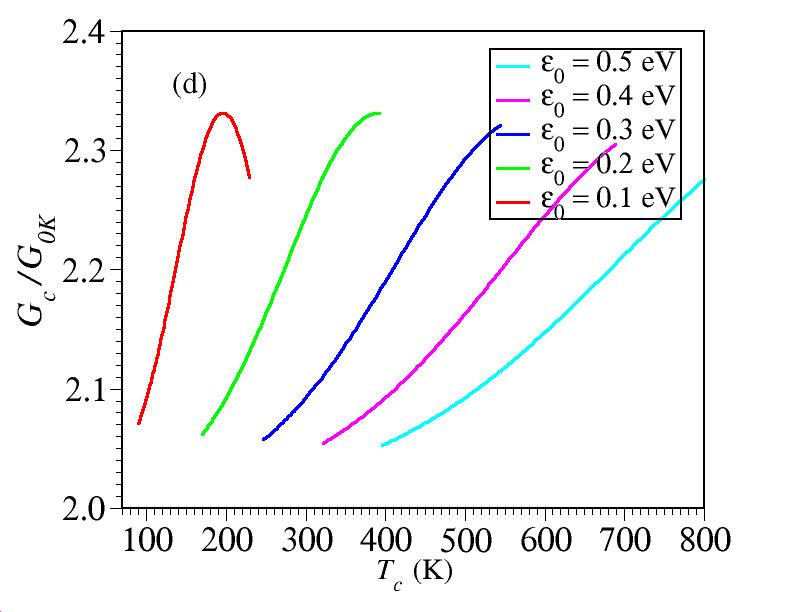}
  }
  \caption{(\textbf{a}) In view of the gradual character of the Arrhenius-Sommerfeld transition, a ``critical'' (``transition'') temperature $T_c$
    can only be defined by an arbitrary convention, e.g., where the inclined dashed lines depicting $G_T$ (\Gl~(\ref{eq-g-sech}))
    cross the cyan horizontal line depicting the zero temperature conductance $G_{0K}$. Notice that at $T = T_c$
    the exact conductance is, roughly,
    two times larger $G_{0K}$ (horizontal magenta line).
    (\textbf{b}) The curve of the critical temperature in dimensionless coordinates obtained using \Gl~(\ref{eq-Tc}). (\textbf{c}) Curves for
    $T_c$ for various model parameter values indicated in the inset. (\textbf{d}) Curves showing that at $T=T_c$ the exact conductance $G_c = G\left(T_c\right)$
    is approximately twice the zero temperature conductance $G_{0K}$.}
  \label{fig:Tc}
\end{figure}

Because this transition is gradual, a crossover (``critical'' or ``transition'') temperature $T_c$
separating these Arrhenius and Sommerfeld regimes can only be defined by some arbitrary convention.
An intuitive possibility is to define $T_c$ by the point where extrapolated (dashed, nearly linear) curves of $G_T$ (\Gl~(\ref{eq-g-sech}))
cross the horizontal (dashed, cyan) line corresponding to the zero temperature value $G_{0K}$, \Gl~(\ref{eq-Tc}) (\figname\ref{fig:Tc}a).
This ``critical'' temperature $T_c$ approximately corresponds to the temperature where the exact, temperature
dependent value of $G$ represents twice the zero temperature value $G_{0K}$ (\Gl~(\ref{eq-Gc}),
magenta horizontal line in \figname\ref{fig:Tc}a,d).
\begin{subequations}
\begin{eqnarray}
  \label{eq-Tc}
  \left . G_T\right\vert_{T=Tc} & = & G_{0K} \\
  \label{eq-Gc}
  \left . G(T)\right\vert_{T=Tc} & \approx & 2 G_{0K}
\end{eqnarray}
\end{subequations}
Imposing \Gl~(\ref{eq-Tc}) yields a curve of $T_c$ versus $\Gamma_a$ which is unique
in the reduced quantities
$T_c/\left\vert \varepsilon_0\right\vert$ and $\Gamma_a/\left\vert \varepsilon_0\right\vert$
(cf.~\figname\ref{fig:Tc}b). More specific illustrations are depicted in \figname\ref{fig:Tc}c,
which give a flavor on the values characterizing real molecular junctions. Noteworthily, the results
presented in \figname\ref{fig:Tc} give additional support to a previous conclusion \cite{Baldea:2017d};
contradicting a possible naive expectation, the crossover between a temperature
dependent and temperature independent transport by tunneling occurs at a value
of $k_B T_c$ which is, in general, substantially different from $\Gamma_a$.
\vspace{-6pt} 
\unskip
\subsection{Results for Specific Molecular Junctions}
\label{sec:real}
Out of the experimental results available for charge transport through molecular junctions at variable temperature
\cite{Poot:06,Reed:09,Zandvliet:12,McCreery:13a,Asadi:13,Tao:16b,McCreery:16a,Nijhuis:16b,Nijhuis:16d,Guo:17,McCreery:17a,Baldea:2018a,Guo:21},
we will consider in this section the junctions fabricated with symmetric molecules
consisting of a ferrocene unit (Fc) \cite{Haaland:68,Coriani:06}
contacted via alkyl spacers to electrodes \cite{Nijhuis:16b} in two testbeds. 
In single molecule transistor (SET) setup,
\ce{-S-(CH2)4-Fc-(CH2)4-S-} molecules were contacted to gold electrodes via thiol groups.
In junctions based on self assembled monolayers (SAM),
molecules were sandwiched between gold and EGaIn electrodes (\ce{Au-S-(CH2)6-Fc-(CH2)6-CH3}/EGaIn).

We compared the theoretical zero bias conductance $G = G(T)$ with the quantity $j(T; V)/V$ estimated from
the experimental currents $j(T; V)$ given in arbitrary units in ref.~\cite{Nijhuis:16b} for the lowest bias $V$
(namely, at $V=10$\,mV for SET for $80\,\mbox{K} \leq T \leq 220$\,K
and at $V=160$\,mV for SAM for $220\,\mbox{K} \leq T \leq 330$\,K).

\begin{figure}[H]
  {\includegraphics[width=0.4\textwidth,angle=0]{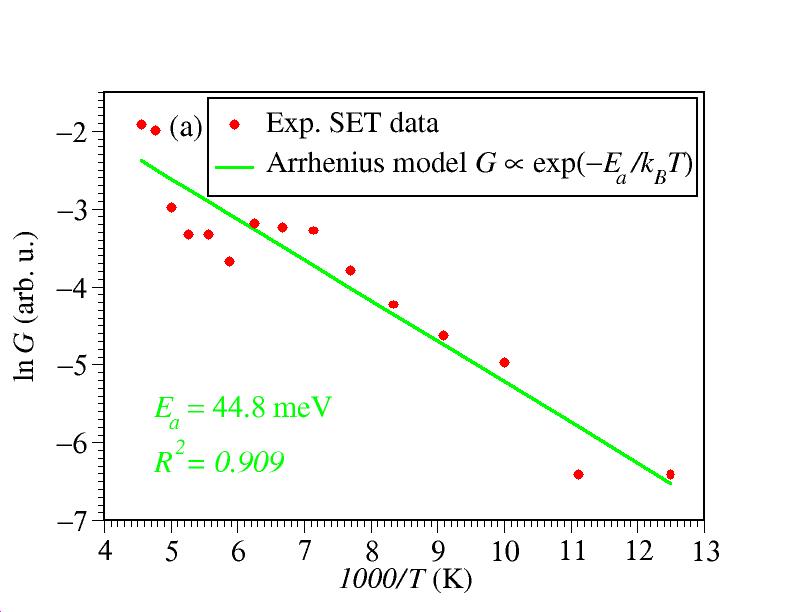}
  \unskip
    \includegraphics[width=0.4\textwidth,angle=0]{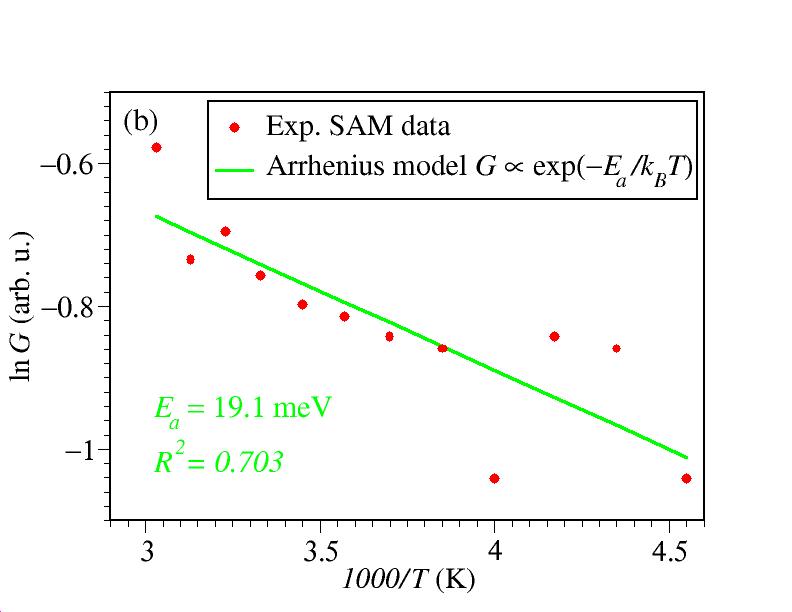}}
   {\includegraphics[width=0.4\textwidth,angle=0]{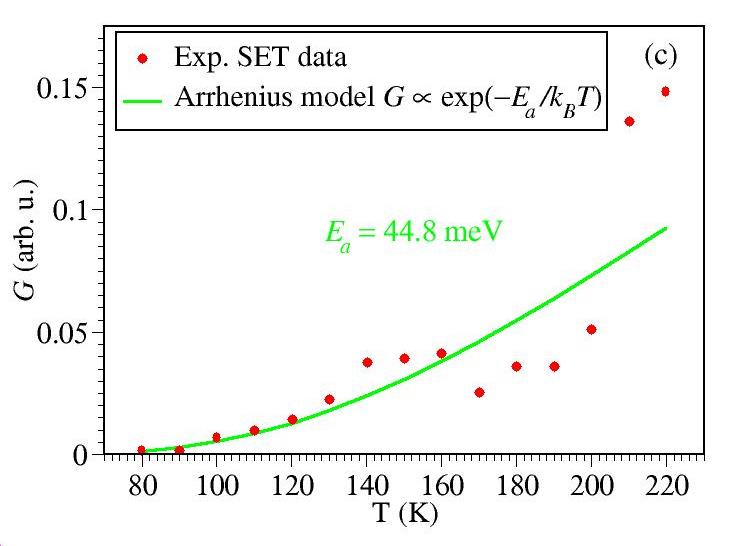}
    \includegraphics[width=0.4\textwidth,angle=0]{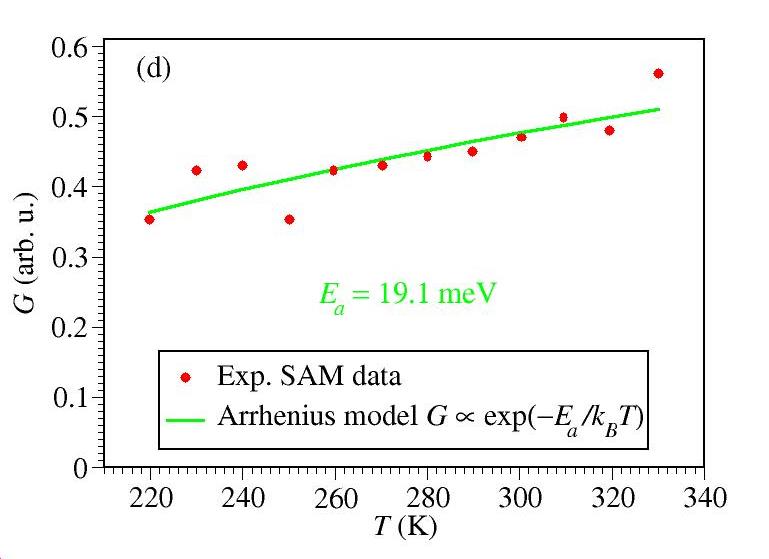}} 
  \caption{{Arrhenius fits }
(\Gl~(\ref{eq-arrhenius})) of the temperature dependent conductance $G$ of the ferrocene-based molecular junctions
    measured in ref.~\cite{Nijhuis:16b} (courtesy of C.~A.~Nijhuis and Y.~Li):
    (\textbf{a}) for SET setup at $V = 10$\,mV and (\textbf{b}) SAM setup at $V= 160$\,mV.
    They are recast in coordinates $G$ versus $T$ in (\textbf{c},\textbf{d}), respectively.
    The values of $V$ indicated here correspond to the lowest bias results reported \cite{Nijhuis:16b}.} 
  \label{fig:arrhenius}
\end{figure}

To start with, we present in \figname\ref{fig:arrhenius} results obtained by fitting the experimental data
(courtesy of C.~A.~Nijhuis and Y.~Li) postulating an Arrhenius dependence
\begin{equation}
  \label{eq-arrhenius}
  G = G_{\infty} \exp\left(-\frac{E_a}{k_B T}\right)
  \end{equation}
which corresponds to $\ln G $ varying linearly with inverse temperature $1/T$. The activation energies $E_a \simeq 45$\,meV for SET and $E_a \simeq 160$\,meV
for SAM obtained using MATHEMATCA's routine {\sl LinearModelFit} shown in \figname\ref{fig:arrhenius}
are similar to those from \figname{3} of ref.~\cite{Nijhuis:16b}.

However, as seen above, a pure Arrhenius dependence cannot be substantiated by the present model calculations.
Model parameters estimated from data fitting using the various methods discussed in \secname~\ref{sec:general}
are collected in Table~\ref{table}. They show that even the
pseudo-Arrhenius form $G \to G_{p.A}$ (\Gl~(\ref{eq-g-exp})), which merely differs from $G_A$ by a prefactor $\propto 1/T$, yields
significantly different ``activation energies'' ($\left\vert\varepsilon_0\right\vert  \simeq 56$\,meV for SET and
$\left\vert\varepsilon_0\right\vert  \simeq 44$\,meV for SAM). We put ``activation energies'' in quotation marks
because $\left\vert\varepsilon_0\right\vert $ does not represent a true barrier energy to be overcome by the charge carriers
(in our specific case of HOMO-mediated conduction, holes \cite{Nijhuis:16b}).
\begin{table}[H]
  \caption{Parameter values estimated using various methods discussed in the main text. All quantities in meV. \label{table}}
\begin{tabular*}{\hsize}{c@{\extracolsep{\fill}}cccc}
\toprule
\textbf{Method} & \textbf{Property} & \textbf{SET}   & \textbf{SAM}\\
\midrule
Equation~(\ref{eq-arrhenius}) & $E_a$                                 &   45  &  19 \\[0.5ex]
Equation~(\ref{eq-g-exp})     & $\left\vert\varepsilon_0\right\vert $ &   56  &  44 \\[0.5ex]
Equation~(\ref{eq-g-sech})    & $\left\vert\varepsilon_0\right\vert $ &   57  &  53 \\[0.5ex]
Equation~(\ref{eq-g-pccp})    & $\left\vert\varepsilon_0\right\vert $ &   58  & 193 \\
                          & $\Gamma_{a}$                          & 0.046 &  11 \\[0.5ex]
Equations~(\ref{eq-g-exact}), (\ref{eq-g-approx-1}) or (\ref{eq-g-approx-2})
                          & $\left\vert\varepsilon_0\right\vert $ &   58  & 238 \\ 
                          & $\Gamma_{a}$                          & 0.039 & 4.6 \\
\bottomrule
\end{tabular*}
\end{table}

We recast the data depicted in Arrhenius coordinates ($\ln G$ versus $1/T$, \figname\ref{fig:arrhenius}a,b)
in coordinates $G$ versus $T$ (\figname\ref{fig:arrhenius}c,d, respectively) to emphasize that,
while not conspicuous for the case of SET,
inferring an Arrhenius dependence from the measurements for the SAM-based junctions is highly problematic.

\begin{figure}[H]
  {
    \includegraphics[width=0.47\textwidth,angle=0]{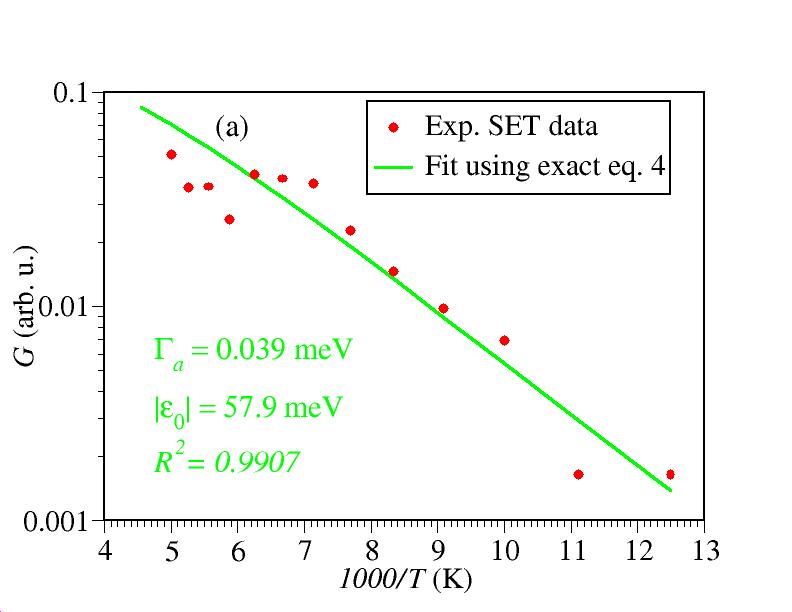}
    \unskip
    \includegraphics[width=0.47\textwidth,angle=0]{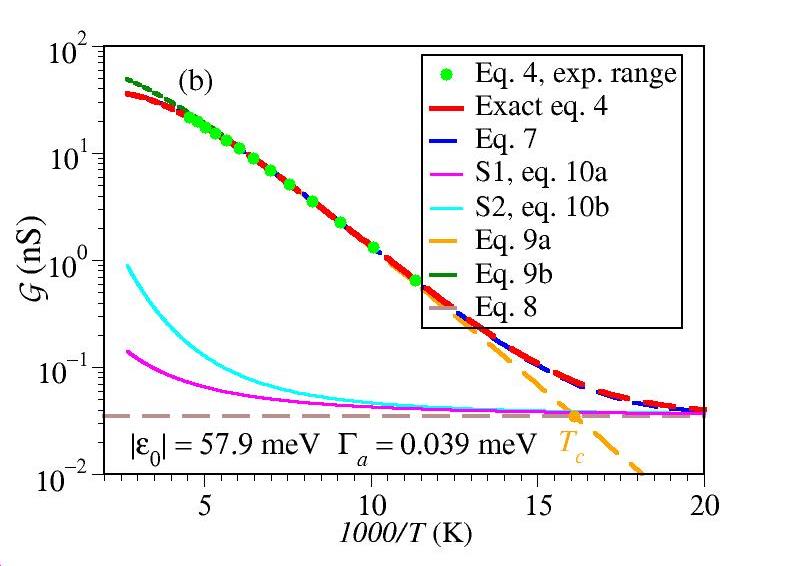}}
  {
    \includegraphics[width=0.47\textwidth,angle=0]{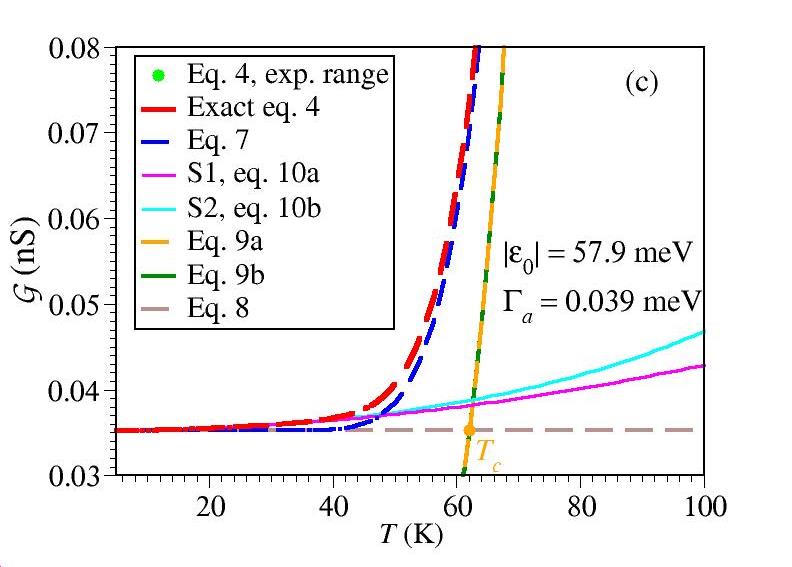}
    \includegraphics[width=0.47\textwidth,angle=0]{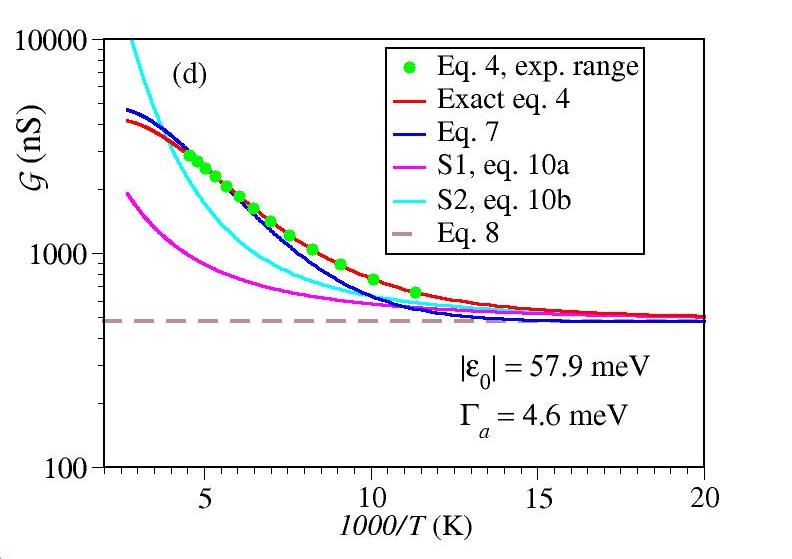}    
  }
  \caption{{Results for SET setup.} 
    (\textbf{a}) Conductance ($G$) data at the lowest bias ($V=10$\,mV) reported in ref.~\cite{Nijhuis:16b} fitted using the exact \Gl~(\ref{eq-g-exact}).
    (\textbf{b}) Exact curve $\mathcal{G}$ versus $1/T$ extrapolated beyond the temperature range sampled experimentally \cite{Nijhuis:16b}
    along with various approximations indicated in the legend.
    (\textbf{c}) Same as (\textbf{b}) recast as a function of $T$.
    (\textbf{d}) Same as (\textbf{b}) using the width value $\Gamma_{a}$ estimated for the SAM setup.
    For the meaning of $\mathcal{G} \propto G$, see \Gl~(\ref{eq-cal-G}) in the main text.
  }
  \label{fig:set}
\end{figure}

\begin{figure}[H]
  {
    \includegraphics[width=0.48\textwidth,angle=0]{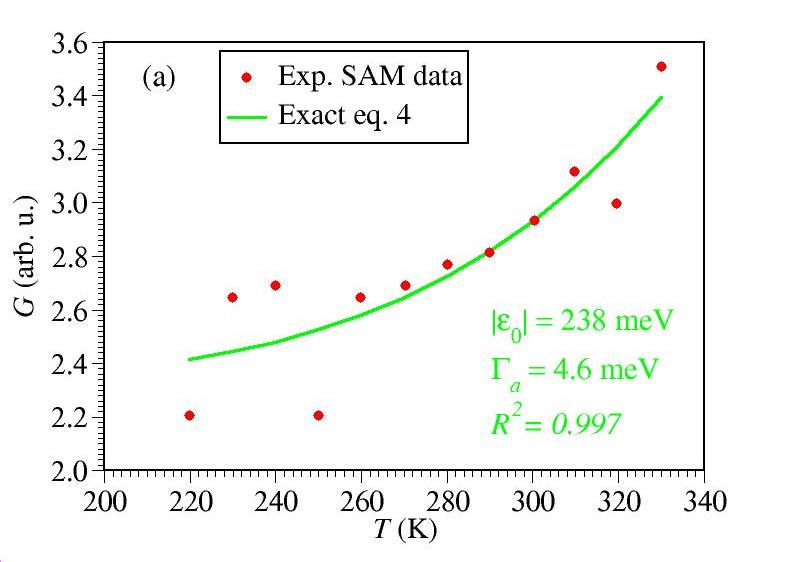}
        \unskip
    \includegraphics[width=0.48\textwidth,angle=0]{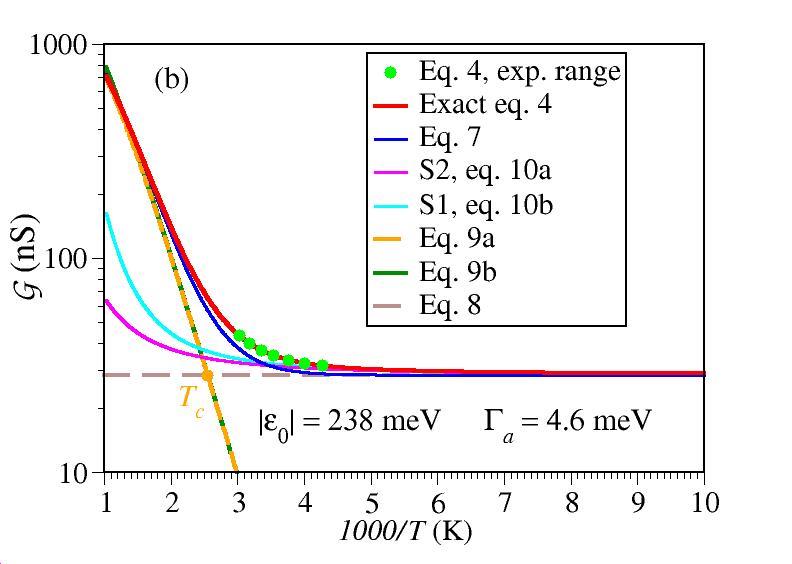}}
  {
    \includegraphics[width=0.48\textwidth,angle=0]{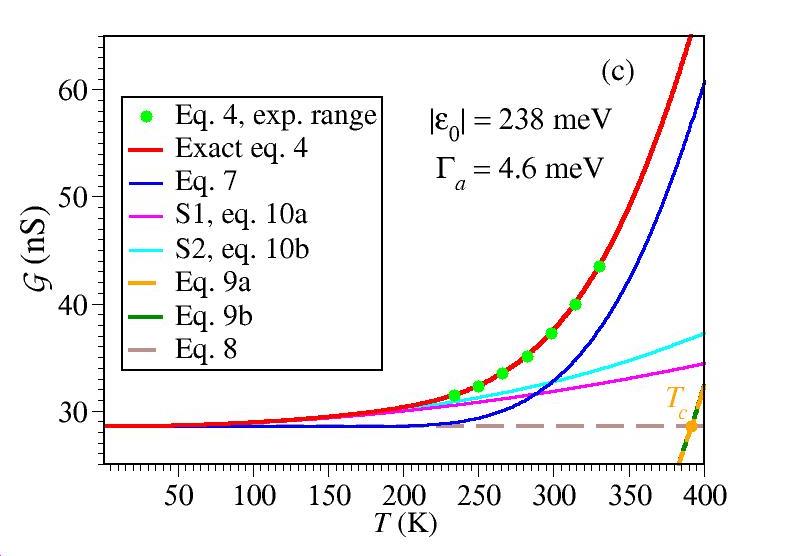}
    \includegraphics[width=0.48\textwidth,angle=0]{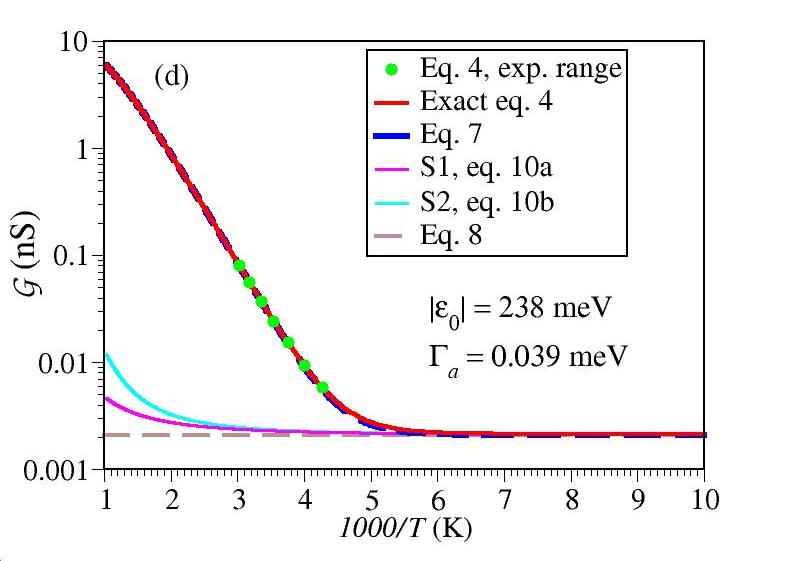}
  }
  \caption{Results for SAM setup.
    (\textbf{a}) Conductance data at the lowest bias ($V=160$\,mV) reported in ref.~\cite{Nijhuis:16b} fitted using the exact \Gl~(\ref{eq-g-exact}).
    (\textbf{b}) Exact curve $\mathcal{G}$ versus $1/T$ extrapolated beyond the temperature range sampled experimentally \cite{Nijhuis:16b}
    along with various approximations indicated in the legend.
    (\textbf{c}) Same as (\textbf{b}) recast as a function of $T$.
    (\textbf{d}) Same as (\textbf{b}) using the width value $\Gamma_{a}$ estimated for the SET setup.
    For the meaning of $\mathcal{G} \propto G$, see \Gl~(\ref{eq-cal-G}) in the main text.}
  \label{fig:sam}
\end{figure}

\Figsname\ref{fig:set}a and \ref{fig:sam}a depict data fitting using the exact \Gl~(\ref{eq-g-exact}) and MATHEMATICA's routine
             {\sl {NonlinearModelFit}. }
Comparison between \figsname\ref{fig:arrhenius}d and \ref{fig:sam}a makes it clear why the MO energy offset 
estimated exactly for SAM ($\left\vert\varepsilon_0\right\vert \simeq 238$\,meV) differs by an order of magnitude from
the Arrhenius-based activation energy ($E_a \simeq 19$\,meV). As visible (and also reflected in the different $R^2$-values),
the fitting curve of \figname\ref{fig:arrhenius}d better describes the general trend emerging the experimental data than the Arrhenius-based
fitting curve of \figname\ref{fig:arrhenius}d.

This difference is not so pronounced in the SET case (cf.~\figsname\ref{fig:arrhenius}a and \ref{fig:set}a).
This explains why, although significant, the difference between the estimated MO energy offset ($\left\vert\varepsilon_0\right\vert \simeq 58$\,meV)
and the Arrhenius-based activation energy ($E_a \simeq 45$\,meV) is not so dramatically large.
\vspace{-9pt} 
%

For comparison purposes, along with the exact curves for conductance,
in \linebreak \mbox{\figsname\ref{fig:set}b,c and \ref{fig:sam}b,c} we also show curves computed with the same
parameters using various approximate formulas presented in \secname\ref{sec:general}.
They are depicted for temperature ranges beyond those (indicated by green points) sampled in experiment,
in order to emphasize that
the experiments of ref.~\cite{Nijhuis:16b} did not sample the Arrhenius-Sommerfeld
transition for SET but partially sampled it for SAM.

\figname\ref{fig:set}b reveals why for SET experiments
\Gl~(\ref{eq-g-pccp}) represents a much more reasonable approximation
than for SAM experiments (\figname\ref{fig:sam}c). In the former case, the temperatures explored experimentally are well below
$T_c$ (the value of which is marked by an orange point), while in the latter case they are above $T_c$.
The small asymptotic (zero temperature) value $G_{0K}$
depicted by the brown dashed line in \figname\ref{fig:set}b makes it clear why
\Gl~(\ref{eq-g-sech},b) still reasonably describe the SET experimental data;
\Gl~(\ref{eq-g-sech}) reasonably approximates \Gl~(\ref{eq-g-pccp}) in cases where $G_{0K}$ is small. 
Again, this is in contrast to the SAM data (\figname\ref{fig:sam}b,c).

Although the temperatures explored experimentally
are above $T_c$, thermal effects exhibited by the SAM data do not merely represent corrections
to the zero temperature limit. The SAM data do not simply belong to the pure Sommerfeld regime;
the magenta (\Gl~(\ref{eq-Sommerfeld-1})) and cyan  (\Gl~(\ref{eq-Sommerfeld-2})) curves in \figname\ref{fig:sam}b,c do significantly
differ from the exact red curve (\Gl~(\ref{eq-g-exact})).
In accord to those elaborated in \secname\ref{sec:general}, one could also note here that
\figname\ref{fig:sam}b,c illustrate limitations of the interpolation expressed by \Gl~(\ref{eq-g-pccp})
in describing the crossover Arrhenius-Sommerfeld regime.

Let us briefly comment on the difference between the parameters for the SET and SAM.
The relatively small difference 
between the values of $\varepsilon_0$ extracted form the SET and SAM data (58\,meV versus 238\,meV, respectively)
can reflect effects due to the gate voltage ($V_g = -1.5$\,V versus $V_g = 0$) \cite{Reed:09,Reed:11} and image charges
(absent in the former case, present in the latter) \cite{Baldea:2013b}.
More importantly than differences in $\varepsilon_0$, $\Gamma_a$'s differ by two orders of magnitude. We assign this difference
as an effect of the SAM-driven work function modification $\delta \Phi$.
The strong (exponential) dependence of the molecule-electrode couplings on $\delta \Phi$ was
amply documented in earlier studies \cite{Frisbie:11,Baldea:2015d,Baldea:2018a,Baldea:2019d,Baldea:2019h}.

To emphasize the important role played by $\Gamma_a$ in the Arrhenius-Sommerfeld transition,
we also show curves for conductance computed for $\varepsilon_0$ determined for the SET setup and
$\Gamma_a$ estimated for the SAM setup (\figname\ref{fig:set}d) and vice versa (\figname\ref{fig:sam}d).
In the former case, the temperature range sampled experimentally comprises the crossover region between the
Arrhenius and Sommerfeld regimes. In the latter case, the temperature range sampled experimentally is shifted 
inside the Arrhenius regime.
\subsection{The Arrhenius-Sommerfeld Thermal Transition:
  A Possible Approach to Estimate the Number of Molecules in Large Area Tunneling Molecular Junctions}
\label{sec:N}
In the various formulas presented above, $G$ is the conductance per molecule.
Therefore, whatever the method utilized, fitting the transport measurements of ref.~\cite{Nijhuis:16b}
encounters an important difficulty: ref.~\cite{Nijhuis:16b} only reported relative currents, not absolute currents.
This is why, paradoxically, the discussion of this specific case is significantly more involved than
the general methodology (\secname\ref{sec:workflow}) to be applied in cases where experimentalists report
absolute (not relative) values of measured currents.

Fitting relative currents using \Gl~(\ref{eq-g-sech},b) 
(as well as \Gl~(\ref{eq-arrhenius}), which was also used in ref.~\cite{Nijhuis:16b}) merely allows the determination of
$\left\vert\varepsilon_0\right\vert$.
Data fitting using \linebreak \mbox{\Gls~(\ref{eq-g-exact}), (\ref{eq-g-approx-1},f) or (\ref{eq-g-pccp})}
allows to obtain the values
of $\left\vert\varepsilon_0\right\vert$ and $\Gamma_a$, but $\Gamma_g$ can only be obtained up to an
unknown multiplicative factor. 

For this reason, the value of $\Gamma_g$ was not indicated in \figsname\ref{fig:set}a and \ref{fig:sam}a,
and $G$ was given in arbitrary units. 
What we showed in \figsname\ref{fig:set}b--d and \ref{fig:sam}b--d is
the conductance per molecule $\mathcal{G}$ defined as
\begin{subequations}
\begin{equation}
  \label{eq-cal-G}
  \mathcal{G} \equiv \left . G\right\vert_{\Gamma_g = \Gamma_a}
\end{equation}
\textls[-25]{which holds when the MO level is symmetrically coupled 
to electrodes (cf.~\mbox{\Gls~(\ref{eq-Gamma-g}) and (\ref{eq-Gamma-a})})}
\begin{equation}
\label{eq-sym-Gamma}  
  \Gamma_a = \Gamma_g \Leftrightarrow \Gamma_s = \Gamma_t
\end{equation}

To exemplify this, and for greater clarity, used in conjunction with \Gl~(\ref{eq-g-exact}), $\mathcal{G}$ is expressed~by
\begin{equation}
  \label{eq-cal-G-exact}
  \frac{\mathcal{G}}{G_0} = \frac{\Gamma_a}{2 \pi k_B T}\,\mbox{\small Re}\, \psi^\prime \left(\frac{1}{2} + \frac{\Gamma_a}{2 \pi k_B T} + i \frac{\varepsilon_0}{2 \pi k_B T}\right)
\end{equation}
\end{subequations}

The assumption in \Gl~(\ref{eq-sym-Gamma}) is justified for the electrostatically gated SET
\linebreak (\ce{Au-S-(CH2)4-Fc-(CH2)4-S-Au})
symmetrically adsorbed chemically via thiol groups, which are very
likely single molecule devices \cite{Reed:09,Baldea:2012c}. For this reason, 
$\mathcal{G}$ presented in \mbox{\figname\ref{fig:set}b} is equal to the true (absolute, i.e., not relative) 
conductance value $G$. The absolute values calculated in this way appear to be consistent with
the absolute values measured in experiment \cite{Nijhuis:16b},
as far as they can be reconstituted after so many years \cite{delBarco}.

Obviously, the above approach cannot be applied for the EGaIn large area SAM-based junctions having a nominal (geometric)
area of $A_{\mbox{\small n} } \approx 700\,\upmu\mbox{m}^2$ \cite{Nijhuis:16b}. The reason is twofold.
First, they comprise an effective number of molecules $N_{\mbox{\small eff}} > 1$.
Above, we said ``nominal area'' and ``effective number'' because, as well documented
\cite{Selzer:05,Milani:07,Boer:08,Whitesides:13,Nijhuis:16g,Frisbie:16d,Cahen:17a,Cahen:20},
the effective (``electric'') area $A_{\mbox{\small eff}}$ may be on orders of magnitude smaller than $A_{\mbox{\small n}}$, or
rephrased, because the total number of molecules $N_{n}$ in the junction is much larger than those effectively
involved in charge transport:
\begin{equation}
  \label{eq-fraction}
  f = \frac{A_{\mbox{\small eff} }}{A_{\mbox{\small n} }} = \frac{N_{\mbox{\small eff} }}{N_{\mbox{\small n} }} \ll 1
\end{equation}
Second, the physical (van der Waals) EGaIn contact with the SAM
is quantified by a coupling $\Gamma_t \equiv \Gamma_{t}^{\mbox{\small EGaIn}}$ substantially smaller than the chemical coupling
$\Gamma_s \equiv \Gamma_{s}^{\mbox{\small Au}}$ to the gold substrate.

Put together, the following relations relating the presently calculated $\mathcal{G}$ and 
the conductance $G_{j}$ of the measured junction apply
\begin{subequations}
  \label{eq-exp}
\begin{eqnarray}
  G & = &  \frac{\Gamma_{t}^{\mbox{\small EGaIn}}}{\Gamma_{s}^{\mbox{\small Au}}} \mathcal{G} \\
  \label{eq-n-eff}
  N_{\mbox{\small eff}} & = & \frac{G_{j}}{G} = \frac{\Gamma_{s}^{\mbox{\small Au}}}{\Gamma_{t}^{\mbox{\small EGaIn}}} \frac{G_{j}}{\mathcal{G}} \\
\label{eq-coverage}
\Sigma & = & \frac{N_{\mbox{\small n}}}{A_{\mbox{\small n}}} =  \frac{N_{\mbox{\small eff}}}{A_{\mbox{\small eff}}}\\
\label{eq-ratio}
\frac{N_{\mbox{\small eff}}}{N_{\mbox{\small n}}} & = & \frac{A_{\mbox{\small eff}}}{A_{\mbox{\small n}}} =
\frac{\Gamma_{s}^{\mbox{\small Au}}}{\Gamma_{t}^{\mbox{\small EGaIn}}}
  \frac{G_{j}}{\mathcal{G}}
  \frac{1}{A_{\mbox{\small n}}\Sigma}
\end{eqnarray}
\end{subequations}
Above, $\Sigma$ stands for the SAM coverage (number of molecules per unit area).

For SAMs of alkyl thiols and oligophenylene thiols utilized to fabricate CP-AFM junctions, measurements via Rutherford backscattering (RBS) and
nuclear reaction analysis (NRA) provided coverage values $\Sigma \simeq 3.5$\,molecules/nm$^2$ practically independent of the type of molecule 
\cite{Frisbie:16e,Baldea:2017e}. 

Experiments have indicated similar coverage values of SAMs anchored via thiols on gold substrate
used to fabricate CP-AFM junctions and EGaIn junctions \cite{Frisbie:16d}. Therefore,
the above value of $\Sigma$ is also reasonable for the presently considered SAM.
For the EGaIn-based junctions of nominal contact area
$A_{\mbox{\small n} } \approx 700\,\upmu\mbox{m}^2$ of ref.~\cite{Nijhuis:16b},
a nominal number of molecules per junction $ N_{\mbox{\small n} } = A_{\mbox{\small n} } \Sigma \approx 2.45 \times 10^6$ can thus be estimated. 

At room temperature, we obtained the value $\mathcal{G} \simeq 37$\,nS.
As far as values measured more than eight years ago can be reconstituted\cite{Nijhuis:16b}, a junctions's conductance $G_{j} \approx 20$\,nS
can be inferred \cite{YuanLi}.
For CP-AFM junctions fabricated with alkyl thiols and gold substrate and tip electrodes, we recently estimated a ratio between the thiol
chemisorbed contact and the methyl physisorbed contact of
\begin{equation}
  \label{eq-mono-vs-di}
  \frac{\Gamma_{s}^{\mbox{\small Au}}}{\Gamma_{t}^{\mbox{\small Au}}} \simeq 37
\end{equation}
If we used these values, we would deduce from \Gl~(\ref{eq-n-eff}) a value $N_{\mbox{\small eff}} \approx 20$,
amounting to $f = N_{\mbox{\small eff}} / N_{\mbox{\small n}} = A_{\mbox{\small eff}} / A_{\mbox{\small n}} \approx 10^{-5}$.
However, for the reason explained below, this value is underestimated.

\Gl~(\ref{eq-mono-vs-di}) assumed that both (substrate and tip/top) electrodes are of gold,
which does not apply to the presently considered Au-(\ldots Fc\ldots)/EGaIn junctions. 
The EGaIn top electrode has a significantly different work function from gold.
Using the dependence on the work function $\Phi$ of the effective coupling for CP-AFM junctions fabricated with
alkyl monothiols (label $m$) and alkyl dithiols (label $d$) \cite{Baldea:2019h}, we deduced
\begin{equation}
  \label{eq-mono-di}
\Gamma_{m,d} \propto e^{\delta_{d,m}\,\Phi}
\end{equation}
where $\delta_m = 1.377\,\mbox{eV}^{-1}$ and $\delta_d = 0.998\,\mbox{eV}^{-1}$. 
Following the method presented in ref.~\cite{Baldea:2021b}, we get
\begin{equation}
  \label{eq-monothiols}
\Gamma_t \propto e^{\left(2\delta_m - \delta_d\right)\, \Phi}
\end{equation}
and this yields
\begin{equation}
  \label{eq-Au-vs-EGaIn}
  \frac{\Gamma_{t}^{\mbox{\small Au}}}{\Gamma_{t}^{\mbox{\small EGaIn}}} \approx
  \exp\left[\left(2\delta_m - \delta_d\right)\left(\Phi_{\mbox{\small Au}} - \Phi_{\mbox{\small EGaIn}}\right)\right] \approx 7
\end{equation}
Above, we used the values
$\Phi_{\mbox{\small EGaIn}} = 4.1$\,eV and $\Phi_{\mbox{\small Au}} = 5.2$\,eV.
The fact that $G \propto \Gamma_g^2 \propto \Gamma_t$ (cf.~\Gl~(\ref{eq-Gamma-g})) translates into a corrected value 
\begin{equation}
  \label{eq-mixed-s-t}
  \frac{\Gamma_{s}^{\mbox{\small Au}}}{\Gamma_{t}^{\mbox{\small EGaIn}}} = 
  \frac{\Gamma_{t}^{\mbox{\small Au}}}{\Gamma_{t}^{\mbox{\small EGaIn}}} \times \frac{\Gamma_{s}^{\mbox{\small Au}}}{\Gamma_{t}^{\mbox{\small Au}}} \approx 260
  \end{equation}
to be used instead of \Gl~(\ref{eq-mono-vs-di}) to compute $f$.
With the above value, \Gl~(\ref{eq-n-eff},c) yield
$N_{\mbox{\small eff}} \approx 140 $ and $A_{\mbox{\small eff}} \approx 40\,\mbox{nm}^2$. Indeed, these values are substantially smaller than
$ N_{\mbox{\small n} } \approx 2.45 \times 10^6 $ and $A_{\mbox{\small n} } \approx 700\,\upmu\mbox{m}^2$ indicated above.
This amounts to
\begin{equation}
  f = N_{\mbox{\small eff}} / N_{\mbox{\small n}} = A_{\mbox{\small eff}} / A_{\mbox{\small n}} \approx 0.6 \times 10^{-4}
\end{equation}
This fraction is comparable with area correction factors obtained using completely different methods reported earlier \cite{Whitesides:14}
for other EGaIn-based junctions. Possibly, this value is a general characteristics of 
the platforms using EGaIn top electrodes.

We have also to mention that oligophenyleneimines (OPI) junctions fabricated using EGaIn/Au electrodes
were claimed \cite{Frisbie:16d} to be 100 times more resistive than OPI Au/Au CP-AFM junctions.
The foregoing analysis found that Fc-based EGaIn junctions with alkyl thiol spacers are (only) seven times
(cf.~\Gl~(\ref{eq-Au-vs-EGaIn})) more resistive than similar CP-AFM junctions.
This suggests that care should be taken when
comparing conducting properties of EGaIn and CP-AFM junctions fabricated with different molecular species, e.g.,
One should distinguish between localized electrons contributing to the dominant (HO)MO (read Fc-based junctions of ref.~\cite{Nijhuis:16b}) and
delocalized electrons (read OPI-based junctions of ref.~\cite{Frisbie:16d}).

\subsection{Workflow for Data Fitting}
\label{sec:workflow}
\begin{figure}[H]
{\includegraphics[width=0.9\textwidth,angle=0]{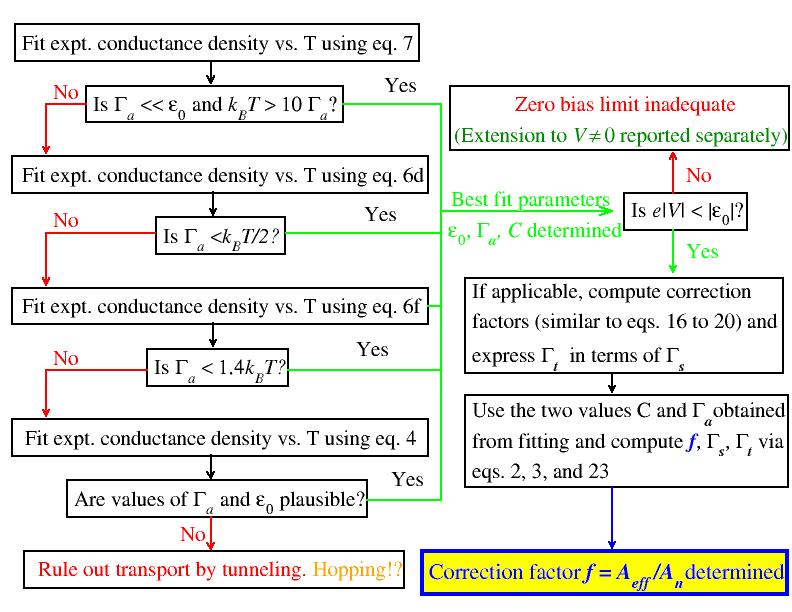}}
    \caption{Diagram depicting the workflow for the presently proposed data fitting approach.}
  \label{fig:workflow}
\end{figure}

In the attempt to aid experimentalists in extracting information from low bias conductance measured at variable temperature,
the workflow for the presently proposed data fitting is summarized in the diagram depicted in \figname\ref{fig:workflow}.
In addition, a few more details may be useful.

Experiments for large area junctions typically report current densities $j_{exp} \equiv j_n$, more precisely, current
($I$) values divided by the junction's \emph{{nominal} 
} area $A_n$.
In the present low bias framework, the envisaged experimental quantity is the nominal conductance density
$g_{exp}(T) \equiv g_{n}(T)$. Straightforward manipulation yields 
\begin{equation}
  \label{eq-exp-eff}
  g_{n}(T)  
  \equiv \left . \frac{j_{n}(V; T)}{V}\right\vert_{V\approx 0}
  \equiv \left . \frac{1}{A_n} \frac{I(V; T)}{V}\right\vert_{V\approx 0}
  \equiv \frac{G_{j}(T)}{A_n}
  = \frac{A_{\mbox{\small eff}}}{A_n} \frac{N_{\mbox{\small eff}}}{A_{\mbox{\small eff}}} G(T)
  = \left(f \Sigma \right) G(T) 
\end{equation}
where $\Sigma$ is the SAM coverage.
One should note that, whether data fitting is based on the exact \Gl~(\ref{eq-g-exact})
or the various approximations based on it---namely,
\Gls~(\ref{eq-g-pccp}), (\ref{eq-g-approx-1}) or
(\ref{eq-g-approx-2}){---,} 
the quantity $\Gamma_g^2$ always enters as a multiplicative factor the RHS of all those expression.
Therefore, by stroke of \Gl~(\ref{eq-exp-eff}), one can use the combination
\begin{equation}
  \label{eq-C}
C \equiv f \Sigma \Gamma_{g}^2
\end{equation}
as a unique fitting parameter. Data fitting based on any of the formulas mentioned above yields best fit estimates
for $\varepsilon_0$, $\Gamma_a$, and $C$.
The area correction factor $f$ can be estimated from $C$ and $\Gamma_a$ by stroke of
\Gls~(\ref{eq-Gamma-g}), (\ref{eq-Gamma-a}) and (\ref{eq-C})
via the additional quantities $\Gamma_{s}$ and $\Gamma_{t}$, provided that an additional relationship
between $\Gamma_{s}$ and $\Gamma_{t}$ exists.

For a specific illustration of how this relationship
can be obtained for EGaIn-based large area junctions with alkyl spacers,
see \Gls~(\ref{eq-mono-vs-di})--(\ref{eq-mixed-s-t}). A similar strategy can be adopted in case
of molecules of oligophenyls \cite{Baldea:2019d,Baldea:2021b} and oligoacenes \cite{Frisbie:11}, for which the contact conductance data
($G_C \propto \Gamma_g^2 = \Gamma_s \Gamma_t$) are also available.

The EGaIn-based junctions represent perhaps the most difficult case to handle. For other platforms
(e.g., CP-AFM or crossed-wire \cite{Kushmerick:02,Kushmerick:02b,Beebe:06}) using symmetric molecules symmetrically contacted
to electrodes, the values of $\Gamma_g = \Gamma_a = \Gamma_s = \Gamma_t$ and $f$ can be straightforwardly
be estimated from \Gl~(\ref{eq-C}). Obviously, provided that absolute values of the current (conductance)
are available, there is no problem at all in the case single molecule junctions. There, $N_{\mbox{\small eff}} = N_{n} = 1$,
and $\Gamma_{g,a}$ and $\varepsilon_0$ can be directly obtained from data fitting.
\unskip
\section{Method}
\label{sec:method}
The method utilized in ths paper is based on the general Keldysh formalism \cite{Caroli:71a,Meir:92,HaugJauho,CuevasScheer:17}
applied to the specified molecular junctions considered.
\section{Conclusions}
\label{sec:conclusion}
Routinely, a curve in ``Arrhenius'' plane ($\ln G$ versus $1/T$) which is a straight line is
taken as evidence for charge transport via a two-step hopping mechanism, while
a plot switching from a linear, inclined line to a horizontal line is taken as revealing a transition from
two-step hopping to single-step tunneling conduction \cite{Choi:08,Tao:10},
and a curve having a slope of magnitude progressively decreasing
as $1/T$ increases is claimed to indicate a variable range hopping mechanism \cite{Shklovskii:84}. 

The curves presented in this paper (e.g.,~\figsname\ref{fig:simul-e0} and \ref{fig:simul-Delta})
demonstrate the drastic limitation of the oversimplified view delineated above.
As we showed, all the aforementioned dependencies are fully compatible with a single-step coherent tunneling
conduction. In a sufficiently broad temperature range, any curve $G$ versus $1/T$
computed by assuming a single-step tunneling mechanism switches 
from a roughly exponential shape (Arrhenius-like regime)
at high $T$ to a less and less $T$-dependent Sommerfeld regime \cite{Baldea:2022c} at low~$T$.

Whether only one of these regimes or both of them can be accessed in a real molecular junction depends, e.g.,
on the value of the crossover (``critical'') temperature $T_c$ (cf.~\mbox{\secname\ref{sec:simulations}} and \figname\ref{fig:Tc}),
on the temperature range that can be sampled experimentally, or on
the thermal stability of the active molecule or electrodes. The latter are significant,
e.g., for protein-based and EGaIn-based junctions, which can be employed in a rather restricted temperature range.

The Arrhenius-Sommerfeld transition 
can be more or less gradual. This is basically controlled by two parameters (level broadening $\Gamma_a$ and $\varepsilon_0$),
which also set the value of $T_c$. 
Unlike $\varepsilon_0$ and $\Gamma_a$, $\Gamma_g$ essentially determines the magnitude of $G$; $T_c$ does not depend on $\Gamma_g$.
In situations far apart from symmetry (e.g., $\Gamma_s \ll \Gamma_t \to \Gamma_g \ll \Gamma_a$),
$\Gamma_g$  only indirectly affects the aforementioned interplay, in the sense that, if $G$ is too small at some temperatures,
the corresponding $T$-range is experimentally irrelevant.

The various theoretical formulas, expressed in closed analytic forms, reported in this paper aims at assisting experimentalists
in processing transport data measured at variable~temperatures.

As an important application of those formulas, we used experimental data for ferrocene-based molecular junctions with
an EGaIn top electrode to illustrate the possibility of estimating the number of molecules per junction,
which is a property of paramount importance for large area junctions, wherein the effective (``electric'') area  $A_{\mbox{\small eff}}$
can and does drastically differ from the nominal (``geometric'') area $A_n$.
For the specific junctions considered, we obtained a value $A_{\mbox{\small eff}}/A_n \approx 0.6 \times 10^{-4}$
compatible with other estimates for EGaIn-based junctions \cite{Whitesides:14}.
To facilitate understanding practical details in implementing
the presently proposed method of estimating the ratio $A_{\text{eff}}/A_n$, we showed a workflow diagram in \figname\ref{fig:workflow}.

In this context, the advantage of the present formulas for $G = G(T)$---\Gls~(\ref{eq-g-exact}), (\ref{eq-g-approx-1},f) and
(\ref{eq-g-pccp})---as compared to other Arrhenius flavors (\Gls~(\ref{eq-g-sech},b) and (\ref{eq-arrhenius}))
used previously in the literature becomes more evident. What the latter formulas can provide is merely an ``activation energy''
whose physical content is more or less obscure. In those formulas, $N \Gamma_g^2 \to N_{\mbox{\small eff}}\Gamma_g^2$
enters as a unique fitting parameter.
From the best fit estimate, $\Gamma_g$ can be computed only in situations where the
effective number of molecules per junction $N_{\text{eff}}$ is known, but it is just this quantity that is the most problematic in case
of large area junctions. For a similar reason, $N_{\text{eff}}/N_n$ 
cannot be confidently determined for cases where the experimentally 
accessed $T$-range merely lies in the nearly exponential fall-off (Arrhenius-like) part of the $G$-curve;
the very weak dependence of $G$ on $\Gamma_a$ in such situations makes even the estimate for $\Gamma_a$
unreliable.

Fortunately, this was not an impediment
in the case of SET \cite{Nijhuis:16b} examined in \secname\ref{sec:real};
although all measured data belong to the Arrhenius regime, the common value of $\Gamma_a \approx \Gamma_g$
can be estimated for symmetric, single-molecule platforms.
With regard to the other (SAM-based) platform considered, the complete characterization of the SAM-based Fc junctions
presented in \secname\ref{sec:real} was possible just because the temperature range explore experimentally
overlaps the Arrhenius-Sommerfeld crossover regime.

To end, we note that the determination of the number of molecules is an important issue not only for large area junctions but also, e.g.,
for CF-AFM junctions.
Although models of contact mechanics \cite{Maugis:92,Johnson:85,Haugstad:12} can be very useful to estimate the
number of molecules in CP-AFM junctions \cite{Baldea:2017e,Baldea:2019d,Baldea:2019h}, reliable information needed
(e.g., values of SAM's Young moduli \cite{Baldea:2021a}) is often missing.
The present method to estimate $N_{\mbox{\small eff}}$ can also applied for the CP-AFM platform.

\ib{Finally, we emphasize that the entire analysis elaborated in the present paper refers to the
transport by tunneling; a coherent, single-step mechanism wherein (say,) electron (or hole) transfer from the left electrode to
the (active) molecule is a process that cannot be separated from the electron transfer from the molecule to the
right electrode. We did not consider the interplay between transport via tunneling and transport via hopping, which is a
two-step mechanism wherein electron transfer from the left electrode to
the molecule and electron transfer from the molecule to the
right electrode are two distinct, uncorrelated processes characterized by durations much shorter than the electron's residence
time on the molecule, which is sufficiently long to allow molecular reorganization \cite{Baldea:2013d}. A possible protocol to disentangle between
tunneling and hopping conduction has been proposed \cite{Baldea:2017d} and applied \cite{Baldea:2018a} elsewhere.}

\vspace{6pt} 
\funding{Financial support from the German Research Foundation
(DFG Grant No. BA 1799/3-2) in the initial stage of this work and computational support by the
state of Baden-W\"urttemberg through bwHPC and the German Research Foundation through
Grant No.~INST 40/575-1 FUGG (bwUniCluster 2, bwForCluster/MLS\&WISO 2/HELIX, and JUSTUS 2 cluster) are gratefully acknowledged. 
}

\acknowledgments{The author thanks Chris Nijhuis and Li Yuan for providing him the raw $I$-$V$-data depicted in
\figname\ref{fig:arrhenius}.
}

\begin{adjustwidth}{-\extralength}{0cm}
\reftitle{References}

\end{adjustwidth}

\begin{thebibliography}{999}

\bibitem[Salomon \em{et~al.}(2003)Salomon, Cahen, Lindsay, Tomfohr, Engelkes,
  and Frisbie]{Frisbie:03}
Salomon, A.; Cahen, D.; Lindsay, S.; Tomfohr, J.; Engelkes, V.; Frisbie, C.
\newblock Comparison of Electronic Transport Measurements on Organic Molecules.
\newblock {\em Adv. Mater.} {\bf 2003}, {\em 15},~1881--1890.
\newblock
  {\changeurlcolor{black}\href{https://doi.org/10.1002/adma.200306091}{\detokenize{https://doi.org/10.1002/adma.200306091}}}.

\bibitem[McCreery and Bergren(2009)]{McCreery:09}
McCreery, R.L.; Bergren, A.J.
\newblock Progress with Molecular Electronic Junctions: Meeting Experimental
  Challenges in Design and Fabrication.
\newblock {\em Adv. Mater.} {\bf 2009}, {\em 21},~4303--4322.
\newblock
  {\changeurlcolor{black}\href{https://doi.org/10.1002/adma.200802850}{{https://doi.org/10.1002/adma.200802850}}}.

\bibitem[McCreery \em{et~al.}(2013)McCreery, Yan, and Bergren]{McCreery:13a}
McCreery, R.L.; Yan, H.; Bergren, A.J.
\newblock A critical perspective on molecular electronic junctions: There is
  plenty of room in the middle.
\newblock {\em Phys. Chem. Chem. Phys.} {\bf 2013}, {\em 15},~1065--1081.
\newblock
  {\changeurlcolor{black}\href{https://doi.org/10.1039/C2CP43516K}{{https://doi.org/10.1039/C2CP43516K}}}.

\bibitem[Xiang \em{et~al.}(2016)Xiang, Wang, Jia, Lee, and Guo]{Guo:16b}
Xiang, D.; Wang, X.; Jia, C.; Lee, T.; Guo, X.
\newblock {Molecular-Scale }
Electronics: From Concept to Function.
\newblock {\em Chem. Rev.} {\bf 2016}, {\em 116},~4318--4440.
  \href{http://xxx.lanl.gov/abs/http://dx.doi.org/10.1021/acs.chemrev.5b00680}
  {\changeurlcolor{black}\href{https://doi.org/10.1021/acs.chemrev.5b00680}{{https://doi.org/10.1021/acs.chemrev.5b00680}}}.

\bibitem[Sangeeth \em{et~al.}(2016)Sangeeth, Demissie, Yuan, Wang, Frisbie, and
  Nijhuis]{Frisbie:16d}
Sangeeth, C.S.S.; Demissie, A.T.; Yuan, L.; Wang, T.; Frisbie, C.D.; Nijhuis,
  C.A.
\newblock Comparison of DC and AC Transport in 1.5--7.5 nm Oligophenylene Imine
  Molecular Wires across Two Junction Platforms: Eutectic Ga-In versus
  Conducting Probe Atomic Force Microscope Junctions.
\newblock {\em J. Am. Chem. Soc.} {\bf 2016}, {\em 138},~7305--7314.
  \href{http://xxx.lanl.gov/abs/http://dx.doi.org/10.1021/jacs.6b02039}
  {\changeurlcolor{black}\href{https://doi.org/10.1021/jacs.6b02039}{{https://doi.org/10.1021/jacs.6b02039}}}.

\bibitem[Mukhopadhyay \em{et~al.}(2020)Mukhopadhyay, Karuppannan, Guo, Fereiro,
  Bergren, Mukundan, Qiu, {Castaneda Ocampo}, Chen, Chiechi, McCreery, Pecht,
  Sheves, Pasula, Lim, Nijhuis, Vilan, and Cahen]{Cahen:20}
Mukhopadhyay, S.; Karuppannan, S.K.; Guo, C.; Fereiro, J.A.; Bergren, A.;
  Mukundan, V.; Qiu, X.; {Castaneda Ocampo}, O.E.; Chen, X.; Chiechi, R.C.;
  et~al.
\newblock Solid-State Protein Junctions: Cross-Laboratory Study Shows
  Preservation of Mechanism at Varying Electronic Coupling.
\newblock {\em iScience} {\bf 2020}, {\em 23},~101099.
\newblock
  {\changeurlcolor{black}\href{https://doi.org/https://doi.org/10.1016/j.isci.2020.101099}{\detokenize{https://doi.org/10.1016/j.isci.2020.101099}}}.

\bibitem[Reed \em{et~al.}(1997)Reed, Zhou, Muller, Burgin, and Tour]{Reed:97}
Reed, M.A.; Zhou, C.; Muller, C.J.; Burgin, T.P.; Tour, J.M.
\newblock Conductance of a Molecular Junction.
\newblock {\em Science} {\bf 1997}, {\em 278},~252--254.
  \href{http://xxx.lanl.gov/abs/http://www.sciencemag.org/cgi/reprint/278/5336/252.pdf}
\newblock
  {\changeurlcolor{black}\href{https://doi.org/10.1126/science.278.5336.252}{{https://doi.org/10.1126/science.278.5336.252}}}.

\bibitem[L\"ortscher \em{et~al.}(2007)L\"ortscher, Weber, and
  Riel]{Loertscher:07}
L\"ortscher, E.; Weber, H.B.; Riel, H.
\newblock Statistical Approach to Investigating Transport through Single
  Molecules.
\newblock {\em Phys. Rev. Lett.} {\bf 2007}, {\em 98},~176807.
\newblock
  {\changeurlcolor{black}\href{https://doi.org/10.1103/PhysRevLett.98.176807}{{https://doi.org/10.1103/PhysRevLett.98.176807}}}.

\bibitem[Reichert \em{et~al.}(2002)Reichert, Ochs, Beckmann, Weber, Mayor, and
  L\"ohneysen]{Reichert:02}
Reichert, J.; Ochs, R.; Beckmann, D.; Weber, H.B.; Mayor, M.; L\"ohneysen, H.V.
\newblock Driving Current through Single Organic Molecules.
\newblock {\em Phys. Rev. Lett.} {\bf 2002}, {\em 88},~176804.
\newblock
  {\changeurlcolor{black}\href{https://doi.org/10.1103/PhysRevLett.88.176804}{{https://doi.org/10.1103/PhysRevLett.88.176804}}}.

\bibitem[Xu and Tao(2003)]{Tao:03}
Xu, B.; Tao, N.J.
\newblock Measurement of Single-Molecule Resistance by Repeated Formation of
  Molecular Junctions.
\newblock {\em Science} {\bf 2003}, {\em 301},~1221--1223.
  \href{http://xxx.lanl.gov/abs/http://www.sciencemag.org/content/301/5637/1221.full.pdf}
\newblock
  {\changeurlcolor{black}\href{https://doi.org/10.1126/science.1087481}{{https://doi.org/10.1126/science.1087481}}}.

\bibitem[Venkataraman \em{et~al.}(2006)Venkataraman, Klare, Nuckolls,
  Hybertsen, and Steigerwald]{Venkataraman:06}
Venkataraman, L.; Klare, J.E.; Nuckolls, C.; Hybertsen, M.S.; Steigerwald, M.L.
\newblock Dependence of Single-Molecule Junction Conductance on Molecular
  Conformation.
\newblock {\em Nature} {\bf 2006}, {\em 442},~904--907.
\newblock
  {\changeurlcolor{black}\href{https://doi.org/http://dx.doi.org/10.1038/nature05037}{\detokenize{http://dx.doi.org/10.1038/nature05037}}}.

\bibitem[Tal \em{et~al.}(2008)Tal, Krieger, Leerink, and van
  Ruitenbeek]{Ruitenbeek:08}
Tal, O.; Krieger, M.; Leerink, B.; van Ruitenbeek, J.M.
\newblock Electron-Vibration Interaction in Single-Molecule Junctions: From
  Contact to Tunneling Regimes.
\newblock {\em Phys. Rev. Lett.} {\bf 2008}, {\em 100},~196804.
\newblock
  {\changeurlcolor{black}\href{https://doi.org/10.1103/PhysRevLett.100.196804}{{https://doi.org/10.1103/PhysRevLett.100.196804}}}.

\bibitem[Song \em{et~al.}(2009)Song, Kim, Jang, Jeong, Reed, and Lee]{Reed:09}
Song, H.; Kim, Y.; Jang, Y.H.; Jeong, H.; Reed, M.A.; Lee, T.
\newblock Observation of Molecular Orbital Gating.
\newblock {\em Nature} {\bf 2009}, {\em 462},~1039--1043.
\newblock
  {\changeurlcolor{black}\href{https://doi.org/10.1038/nature08639}{{https://doi.org/10.1038/nature08639}}}.

\bibitem[Song \em{et~al.}(2011)Song, Reed, and Lee]{Reed:11}
Song, H.; Reed, M.A.; Lee, T.
\newblock Single Molecule Electronic Devices.
\newblock {\em Adv. Mater.} {\bf 2011}, {\em 23},~1583--1608.
\newblock
  {\changeurlcolor{black}\href{https://doi.org/10.1002/adma.201004291}{{https://doi.org/10.1002/\linebreak adma.201004291}}}.

\bibitem[Garrigues \em{et~al.}(2016)Garrigues, Yuan, Wang, Singh, del Barco,
  and Nijhuis]{Nijhuis:16b}
Garrigues, A.R.; Yuan, L.; Wang, L.; Singh, S.; del Barco, E.; Nijhuis, C.A.
\newblock Temperature Dependent Charge Transport across Tunnel Junctions of
  Single-Molecules and Self-Assembled Monolayers: A Comparative Study.
\newblock {\em Dalton Trans.} {\bf 2016}, {\em 45},~17153--17159.
\newblock
  {\changeurlcolor{black}\href{https://doi.org/10.1039/C6DT03204D}{{https://doi.org/10.1039/C6DT03204D}}}.

\bibitem[Wold and Frisbie(2000)]{Frisbie:00}
Wold, D.J.; Frisbie, C.D.
\newblock Formation of Metal-Molecule-Metal Tunnel Junctions: Microcontacts to
  Alkanethiol Monolayers with a Conducting AFM Tip.
\newblock {\em J. Am. Chem. Soc.} {\bf 2000}, {\em 122},~2970--2971.
  \href{http://xxx.lanl.gov/abs/http://pubs.acs.org/doi/pdf/10.1021/ja994468h}
\newblock
  {\changeurlcolor{black}\href{https://doi.org/10.1021/ja994468h}{{https://doi.org/10.1021/ja994468h}}}.

\bibitem[Wold and Frisbie(2001)]{Frisbie:01}
Wold, D.J.; Frisbie, C.D.
\newblock Fabrication and Characterization of Metal-Molecule-Metal Junctions by
  Conducting Probe Atomic Force Microscopy.
\newblock {\em J. Am. Chem. Soc.} {\bf 2001}, {\em 123},~5549--5556.
  \href{http://xxx.lanl.gov/abs/http://pubs.acs.org/doi/pdf/10.1021/ja0101532}
  {\changeurlcolor{black}\href{https://doi.org/10.1021/ja0101532}{{https://doi.org/10.1021/ja0101532}}}.

\bibitem[Beebe \em{et~al.}(2002)Beebe, Engelkes, Miller, and
  Frisbie]{Frisbie:02}
Beebe, J.M.; Engelkes, V.B.; Miller, L.L.; Frisbie, C.D.
\newblock Contact Resistance in Metal-Molecule-Metal Junctions Based on
  Aliphatic SAMs: Effects of Surface Linker and Metal Work Function.
\newblock {\em J. Am. Chem. Soc.} {\bf 2002}, {\em 124},~11268--11269.
  \href{http://xxx.lanl.gov/abs/http://pubs.acs.org/doi/pdf/10.1021/ja0268332}
  {\changeurlcolor{black}\href{https://doi.org/10.1021/ja0268332}{{https://doi.org/10.1021/ja0268332}}}.

\bibitem[Wold \em{et~al.}(2002)Wold, Haag, Rampi, and Frisbie]{Frisbie:02b}
Wold, D.J.; Haag, R.; Rampi, M.A.; Frisbie, C.D.
\newblock Distance Dependence of Electron Tunneling through Self-Assembled
  Monolayers Measured by Conducting Probe Atomic Force Microscopy: Unsaturated
  versus Saturated Molecular Junctions.
\newblock {\em J. Phys. Chem. B} {\bf 2002}, {\em 106},~2813--2816.
  \href{http://xxx.lanl.gov/abs/http://pubs.acs.org/doi/pdf/10.1021/jp013476t}
\newblock
  {\changeurlcolor{black}\href{https://doi.org/10.1021/jp013476t}{{https://doi.org/10.1021/jp013476t}}}.

\bibitem[Engelkes \em{et~al.}(2004)Engelkes, Beebe, and Frisbie]{Frisbie:04}
Engelkes, V.B.; Beebe, J.M.; Frisbie, C.D.
\newblock Length-Dependent Transport in Molecular Junctions Based on SAMs of
  Alkanethiols and Alkanedithiols: Effect of Metal Work Function and Applied
  Bias on Tunneling Efficiency and Contact Resistance.
\newblock {\em J. Am. Chem. Soc.} {\bf 2004}, {\em 126},~14287--14296.
  \href{http://xxx.lanl.gov/abs/http://pubs.acs.org/doi/pdf/10.1021/ja046274u}
\newblock
  {\changeurlcolor{black}\href{https://doi.org/10.1021/ja046274u}{{https://doi.org/10.1021/ja046274u}}}.

\bibitem[Kushmerick \em{et~al.}(2002{\natexlab{a}})Kushmerick, Holt, Pollack,
  Ratner, Yang, Schull, Naciri, Moore, and Shashidhar]{Kushmerick:02}
Kushmerick, J.G.; Holt, D.B.; Pollack, S.K.; Ratner, M.A.; Yang, J.C.; Schull,
  T.L.; Naciri, J.; Moore, M.H.; Shashidhar, R.
\newblock Effect of Bond-Length Alternation in Molecular Wires.
\newblock {\em J. Am. Chem. Soc.} {\bf 2002}, {\em 124},~10654--10655.
  \href{http://xxx.lanl.gov/abs/http://pubs.acs.org/doi/pdf/10.1021/ja027090n}
\newblock
  {\changeurlcolor{black}\href{https://doi.org/10.1021/ja027090n}{{https://doi.org/10.1021/ja027090n}}}.

\bibitem[Kushmerick \em{et~al.}(2002{\natexlab{b}})Kushmerick, Holt, Yang,
  Naciri, Moore, and Shashidhar]{Kushmerick:02b}
Kushmerick, J.G.; Holt, D.B.; Yang, J.C.; Naciri, J.; Moore, M.H.; Shashidhar,
  R.
\newblock Metal-Molecule Contacts and Charge Transport across Monomolecular
  Layers: Measurement and Theory.
\newblock {\em Phys. Rev. Lett.} {\bf 2002}, {\em 89},~086802.
\newblock
  {\changeurlcolor{black}\href{https://doi.org/10.1103/PhysRevLett.89.086802}{{https://doi.org/10.1103/\linebreak PhysRevLett.89.086802}}}.

\bibitem[Kushmerick(2005)]{Kushmerick:05}
Kushmerick, J.G.
\newblock Metal-molecule contacts.
\newblock {\em Mater. Today} {\bf 2005}, {\em 8},~26--30.
\newblock
  {\changeurlcolor{black}\href{https://doi.org/10.1016/S1369-7021(05)70984-6}{{https://doi.org/10.1016/S1369-7021(05)70984-6}}}.

\bibitem[Beebe \em{et~al.}(2006)Beebe, Kim, Gadzuk, Frisbie, and
  Kushmerick]{Beebe:06}
Beebe, J.M.; Kim, B.; Gadzuk, J.W.; Frisbie, C.D.; Kushmerick, J.G.
\newblock Transition from Direct Tunneling to Field Emission in
  Metal-Molecule-Metal Junctions.
\newblock {\em Phys. Rev. Lett.} {\bf 2006}, {\em 97},~026801.
\newblock
  {\changeurlcolor{black}\href{https://doi.org/10.1103/PhysRevLett.97.026801}{{https://doi.org/10.1103/PhysRevLett.97.026801}}}.

\bibitem[Beebe \em{et~al.}(2008)Beebe, Kim, Frisbie, and Kushmerick]{Beebe:08}
Beebe, J.M.; Kim, B.; Frisbie, C.D.; Kushmerick, J.G.
\newblock Measuring Relative Barrier Heights in Molecular Electronic Junctions
  with Transition Voltage Spectroscopy.
\newblock {\em ACS Nano} {\bf 2008}, {\em 2},~827--832.
\newblock
  {\changeurlcolor{black}\href{https://doi.org/10.1021/nn700424u}{{https://doi.org/10.1021/nn700424u}}}.

\bibitem[Simeone \em{et~al.}(2013)Simeone, Yoon, Thuo, Barber, Smith, and
  Whitesides]{Whitesides:13}
Simeone, F.C.; Yoon, H.J.; Thuo, M.M.; Barber, J.R.; Smith, B.; Whitesides,
  G.M.
\newblock Defining the Value of Injection Current and Effective Electrical
  Contact Area for EGaIn-Based Molecular Tunneling Junctions.
\newblock {\em J. Am. Chem. Soc.} {\bf 2013}, {\em 135},~18131--18144.
  \href{http://xxx.lanl.gov/abs/http://dx.doi.org/10.1021/ja408652h}
  {\changeurlcolor{black}\href{https://doi.org/10.1021/ja408652h}{{https://doi.org/10.1021/ja408652h}}}.

\bibitem[Yoon \em{et~al.}(2014)Yoon, Bowers, Baghbanzadeh, and
  Whitesides]{Whitesides:14}
Yoon, H.J.; Bowers, C.M.; Baghbanzadeh, M.; Whitesides, G.M.
\newblock The Rate of Charge Tunneling Is Insensitive to Polar Terminal Groups
  in Self-Assembled Monolayers in AgTSS(CH$_2$)nM(CH$_2$)mT//Ga$_2$O$_3$/EGaIn Junctions.
\newblock {\em J. Am. Chem. Soc.} {\bf 2014}, {\em 136},~16--19.
  \href{http://xxx.lanl.gov/abs/https://doi.org/10.1021/ja409771u}
  {\changeurlcolor{black}\href{https://doi.org/10.1021/ja409771u}{{https://doi.org/10.1021/ja409771u}}}.

\bibitem[Zhao \em{et~al.}()Zhao, Soni, Lee, Nijhuis, and Xiang]{Zhao:22}
Zhao, Z.; Soni, S.; Lee, T.; Nijhuis, C.A.; Xiang, D.
\newblock Smart Eutectic Gallium-Indium: From Properties to Applications.
\newblock {\em Adv. Mater.} 2022, \emph{Early View}.
  \href{http://xxx.lanl.gov/abs/https://onlinelibrary.wiley.com/doi/pdf/10.1002/adma.202203391}
\newblock
  {\changeurlcolor{black}\href{https://doi.org/https://doi.org/10.1002/adma.202203391}{\detokenize{https://doi.org/10.1002/adma.202203391}}}.

\bibitem[Park \em{et~al.}(2019)Park, Kang, and Yoon]{Park:19b}
Park, S.; Kang, S.; Yoon, H.J.
\newblock Power Factor of One Molecule Thick Films and Length Dependence.
\newblock {\em ACS Cent. Sci.} {\bf 2019}, {\em 5},~1975--1982.
  \href{http://xxx.lanl.gov/abs/https://doi.org/10.1021/acscentsci.9b01042}
\newblock
  {\changeurlcolor{black}\href{https://doi.org/10.1021/acscentsci.9b01042}{{https://doi.org/10.1021/acscentsci.9b01042}}}.

\bibitem[Guo \em{et~al.}(2011)Guo, Hihath, Diez-P\'erez, and Tao]{Guo:11}
Guo, S.; Hihath, J.; Diez-P\'erez, I.; Tao, N.
\newblock Measurement and Statistical Analysis of Single-Molecule
  Current-Voltage Characteristics, Transition Voltage Spectroscopy, and
  Tunneling Barrier Height.
\newblock {\em J. Am. Chem. Soc.} {\bf 2011}, {\em 133},~19189--19197.
  \href{http://xxx.lanl.gov/abs/http://pubs.acs.org/doi/pdf/10.1021/ja2076857}
  {\changeurlcolor{black}\href{https://doi.org/10.1021/ja2076857}{{https://doi.org/10.1021/ja2076857}}}.

\bibitem[Li \em{et~al.}(2008)Li, Pobelov, Wandlowski, Bagrets, Arnold, and
  Evers]{Wandlowski:08c}
Li, C.; Pobelov, I.; Wandlowski, T.; Bagrets, A.; Arnold, A.; Evers, F.
\newblock Charge Transport in Single Au|Alkanedithiol|Au Junctions:
  Coordination Geometries and Conformational Degrees of Freedom.
\newblock {\em J. Am. Chem. Soc.} {\bf 2008}, {\em 130},~318--326.
  \href{http://xxx.lanl.gov/abs/http://pubs.acs.org/doi/pdf/10.1021/ja0762386}
\newblock
  {\changeurlcolor{black}\href{https://doi.org/10.1021/ja0762386}{{https://doi.org/10.1021/ja0762386}}}.

\bibitem[Kim \em{et~al.}(2011)Kim, Choi, Zhu, and Frisbie]{Frisbie:11}
Kim, B.; Choi, S.H.; Zhu, X.Y.; Frisbie, C.D.
\newblock Molecular Tunnel Junctions Based on $\pi$-Conjugated Oligoacene
  Thiols and Dithiols between Ag, Au, and Pt Contacts: Effect of Surface
  Linking Group and Metal Work Function.
\newblock {\em J. Am. Chem. Soc.} {\bf 2011}, {\em 133},~19864--19877.
  \href{http://xxx.lanl.gov/abs/http://pubs.acs.org/doi/pdf/10.1021/ja207751w}
\newblock
  {\changeurlcolor{black}\href{https://doi.org/10.1021/ja207751w}{{https://doi.org/10.1021/ja207751w}}}.

\bibitem[Thuo \em{et~al.}(2011)Thuo, Reus, Nijhuis, Barber, Kim, Schulz, and
  Whitesides]{Whitesides:11}
Thuo, M.M.; Reus, W.F.; Nijhuis, C.A.; Barber, J.R.; Kim, C.; Schulz, M.D.;
  Whitesides, G.M.
\newblock Odd-Even Effects in Charge Transport across Self-Assembled
  Monolayers.
\newblock {\em J. Am. Chem. Soc.} {\bf 2011}, {\em 133},~2962--2975.
  \href{http://xxx.lanl.gov/abs/http://dx.doi.org/10.1021/ja1090436}
  {\changeurlcolor{black}\href{https://doi.org/10.1021/ja1090436}{{https://doi.org/10.1021/ja1090436}}}.

\bibitem[Ramin and Jabbarzadeh(2011)]{Ramin:11}
Ramin, L.; Jabbarzadeh, A.
\newblock Odd--Even Effects on the Structure, Stability, and Phase Transition
  of Alkanethiol Self-Assembled Monolayers.
\newblock {\em Langmuir} {\bf 2011}, {\em 27},~9748--9759.
  \href{http://xxx.lanl.gov/abs/https://doi.org/10.1021/la201467b}
  {\changeurlcolor{black}\href{https://doi.org/10.1021/la201467b}{{https://doi.org/10.1021/la201467b}}}.

\bibitem[Baghbanzadeh \em{et~al.}(2014)Baghbanzadeh, Simeone, Bowers, Liao,
  Thuo, Baghbanzadeh, Miller, Carmichael, and Whitesides]{Whitesides:14c}
Baghbanzadeh, M.; Simeone, F.C.; Bowers, C.M.; Liao, K.C.; Thuo, M.;
  Baghbanzadeh, M.; Miller, M.S.; Carmichael, T.B.; Whitesides, G.M.
\newblock Odd-Even Effects in Charge Transport across n-Alkanethiolate-Based
  SAMs.
\newblock {\em J. Am. Chem. Soc.} {\bf 2014}, {\em 136},~16919--16925.
  \href{http://xxx.lanl.gov/abs/https://doi.org/10.1021/ja509436k}
  {\changeurlcolor{black}\href{https://doi.org/10.1021/ja509436k}{{https://doi.org/10.1021/ja509436k}}}.

\bibitem[Jiang \em{et~al.}(0)Jiang, Sangeeth, and Nijhuis]{Nijhuis:15b}
Jiang, L.; Sangeeth, C.S.S.; Nijhuis, C.A.
\newblock The Origin of the Odd-Even Effect in the Tunneling Rates across EGaIn
  Junctions with Self-Assembled Monolayers (SAMs) of n-Alkanethiolates.
\newblock {\em J. Am. Chem. Soc.} {\bf 2015}, {\em 137},~10659--10667.
  \href{http://xxx.lanl.gov/abs/https://dx.doi.org/10.1021/jacs.5b05761}
  {\changeurlcolor{black}\href{https://doi.org/10.1021/jacs.5b05761}{{https://doi.org/10.1021/jacs.5b05761}}}.

\bibitem[Nurbawono \em{et~al.}(2015)Nurbawono, Liu, Nijhuis, and
  Zhang]{Nijhuis:15f}
Nurbawono, A.; Liu, S.; Nijhuis, C.A.; Zhang, C.
\newblock Odd-Even Effects in Charge Transport through Self-Assembled Monolayer
  of Alkanethiolates.
\newblock {\em J. Phys. Chem. C} {\bf 2015}, {\em 119},~5657--5662.
  \href{http://xxx.lanl.gov/abs/https://doi.org/10.1021/jp5116146}
\newblock
  {\changeurlcolor{black}\href{https://doi.org/10.1021/jp5116146}{{https://doi.org/10.1021/jp5116146}}}.

\bibitem[Song \em{et~al.}(2017)Song, Thompson, Annadata, Guerin, Loh, and
  Nijhuis]{Nijhuis:17c}
Song, P.; Thompson, D.; Annadata, H.V.; Guerin, S.; Loh, K.P.; Nijhuis, C.A.
\newblock Supramolecular Structure of the Monolayer Triggers Odd-Even Effects
  in the Tunneling Rates across Noncovalent Junctions on Graphene.
\newblock {\em J. Phys. Chem. C} {\bf 2017}, {\em 121},~4172--4180.
  \href{http://xxx.lanl.gov/abs/https://doi.org/10.1021/acs.jpcc.6b12949}
\newblock
  {\changeurlcolor{black}\href{https://doi.org/10.1021/acs.jpcc.6b12949}{{https://doi.org/10.1021/acs.jpcc.6b12949}}}.

\bibitem[Ben~Amara \em{et~al.}(2020)Ben~Amara, Dionne, Kassir, Pellerin, and
  Badia]{BenAmara:20}
Ben~Amara, F.; Dionne, E.R.; Kassir, S.; Pellerin, C.; Badia, A.
\newblock Molecular Origin of the Odd-Even Effect of Macroscopic Properties of
  n-Alkanethiolate Self-Assembled Monolayers: Bulk or Interface?
\newblock {\em J. Am. Chem. Soc.} {\bf 2020}, {\em
  142},~13051--13061.
  \href{http://xxx.lanl.gov/abs/https://doi.org/10.1021/jacs.0c04288}
  {\changeurlcolor{black}\href{https://doi.org/10.1021/jacs.0c04288}{{https://doi.org/10.1021/jacs.0c04288}}}.

\bibitem[Selzer \em{et~al.}(2005)Selzer, Cai, Cabassi, Yao, Tour, Mayer, and
  Allara]{Selzer:05}
Selzer, Y.; Cai, L.; Cabassi, M.A.; Yao, Y.; Tour, J.M.; Mayer, T.S.; Allara,
  D.L.
\newblock Effect of Local Environment on Molecular Conduction: Isolated
  Molecule versus Self-Assembled Monolayer.
\newblock {\em Nano Lett.} {\bf 2005}, {\em 5},~61--65.
  \href{http://xxx.lanl.gov/abs/https://doi.org/10.1021/nl048372j}
  {\changeurlcolor{black}\href{https://doi.org/10.1021/nl048372j}{{https://doi.org/10.1021/nl048372j}}}.

\bibitem[Milani \em{et~al.}(2007)Milani, Grave, Ferri, Samori, and
  Rampi]{Milani:07}
Milani, F.; Grave, C.; Ferri, V.; Samori, P.; Rampi, M.A.
\newblock Ultrathin $\pi$-Conjugated Polymer Films for Simple Fabrication of
  Large-Area Molecular Junctions.
\newblock {\em ChemPhysChem} {\bf 2007}, {\em 8},~515--518.
  \href{http://xxx.lanl.gov/abs/https://chemistry-europe.onlinelibrary.wiley.com/doi/pdf/10.1002/cphc.200600672}
\newblock
  {\changeurlcolor{black}\href{https://doi.org/https://doi.org/10.1002/cphc.200600672}{\detokenize{https://doi.org/10.1002/cphc.200600672}}}.

\bibitem[Akkerman and de~Boer(2008)]{Boer:08}
Akkerman, H.B.; de~Boer, B.
\newblock Electrical Conduction through Single Molecules and Self-Assembled
  Monolayers.
\newblock {\em J. Phys. Condens. Matt.} {\bf 2008}, {\em 20},~013001.

\bibitem[Suchand~Sangeeth \em{et~al.}(2015)Suchand~Sangeeth, Wan, and
  Nijhuis]{Nijhuis:16g}
Suchand~Sangeeth, C.S.; Wan, A.; Nijhuis, C.A.
\newblock Probing the nature and resistance of the molecule-electrode contact
  in SAM-based junctions.
\newblock {\em Nanoscale} {\bf 2015}, {\em 7},~12061--12067.
\newblock
  {\changeurlcolor{black}\href{https://doi.org/10.1039/C5NR02570B}{{https://doi.org/10.1039/C5NR02570B}}}.

\bibitem[Vilan \em{et~al.}(2017)Vilan, Aswal, and Cahen]{Cahen:17a}
Vilan, A.; Aswal, D.; Cahen, D.
\newblock Large-Area, Ensemble Molecular Electronics: Motivation and
  Challenges.
\newblock {\em Chem. Rev.} {\bf 2017}, {\em 117},~4248--4286.
  \href{http://xxx.lanl.gov/abs/http://dx.doi.org/10.1021/acs.chemrev.6b00595}
  {\changeurlcolor{black}\href{https://doi.org/10.1021/acs.chemrev.6b00595}{{https://doi.org/10.1021/acs.chemrev.6b00595}}}.

\bibitem[B\^aldea(2022)]{Baldea:2022c}
B\^aldea, I.
\newblock Exact Analytic Formula for Conductance Predicting a Tunable
  Sommerfeld-Arrhenius Thermal Transition within a Single-Step Tunneling
  Mechanism in Molecular Junctions Subject to Mechanical Stretching.
\newblock {\em Adv. Theor. Simul.} {\bf 2022}, {\em 5},~202200158.
  \href{http://xxx.lanl.gov/abs/https://onlinelibrary.wiley.com/doi/pdf/10.1002/adts.202200158}
\newblock
  {\changeurlcolor{black}\href{https://doi.org/https://doi.org/10.1002/adts.202200158}{\detokenize{https://doi.org/10.1002/adts.202200158}}}.

\bibitem[Caroli \em{et~al.}(1971)Caroli, Combescot, Nozieres, and
  Saint-James]{Caroli:71a}
Caroli, C.; Combescot, R.; Nozieres, P.; Saint-James, D.
\newblock Direct Calculation of the Tunneling Current.
\newblock {\em J. Phys. C Solid State Phys.} {\bf 1971}, {\em 4},~916.
\newblock
  {\changeurlcolor{black}\href{https://doi.org/10.1088/0022-3719/4/8/018}{{https://doi.org/10.1088/0022-3719/4/8/018}}}.

\bibitem[Meir and Wingreen(1992)]{Meir:92}
Meir, Y.; Wingreen, N.S.
\newblock Landauer formula for the current through an interacting electron
  region.
\newblock {\em Phys. Rev. Lett.} {\bf 1992}, {\em 68},~2512--2515.
\newblock
  {\changeurlcolor{black}\href{https://doi.org/10.1103/PhysRevLett.68.2512}{{https://doi.org/10.1103/PhysRevLett.68.2512}}}.

\bibitem[Haug and Jauho(2008)]{HaugJauho}
Haug, H.J.W.; Jauho, A.P.
\newblock {\em Quantum Kinetics in Transport and Optics of Semiconductors},
  2nd ed.; Springer Series in Solid-State
  Sciences: Berlin/Heidelberg, Germany; New York, NY, USA, 2008; Volume 123. 
\newblock
  {\changeurlcolor{black}\href{https://doi.org/DOI:10.1007/978-3-540-73564-9}{\detokenize{https://doi.org/10.1007/978-3-540-73564-9}}}.

\bibitem[Cuevas and Scheer(2017)]{CuevasScheer:17}
Cuevas, J.C.; Scheer, E.
\newblock {\em Molecular Electronics: An Introduction to Theory and
  Experiment}, 2nd ed.; World Scientific Series in Nanoscience and Nanotechnology; World Scientific: {London, UK,} 
 2017; Volume 15.
  \href{http://xxx.lanl.gov/abs/https://www.worldscientific.com/doi/pdf/10.1142/10598}
  {\changeurlcolor{black}\href{https://doi.org/10.1142/10598}{{https://doi.org/10.1142/10598}}}.

\bibitem[B\^aldea(2017)]{Baldea:2017d}
B\^aldea, I.
\newblock Protocol for Disentangling the Thermally Activated Contribution to
  the Tunneling-Assisted Charge Transport. Analytical Results and Experimental
  Relevance.
\newblock {\em Phys. Chem. Chem. Phys.} {\bf 2017}, {\em 19},~11759--11770.
\newblock
  {\changeurlcolor{black}\href{https://doi.org/10.1039/C7CP01103B}{{https://doi.org/10.1039/C7CP01103B}}}.

\bibitem[B\^aldea(2021)]{Baldea:2021b}
B\^aldea, I.
\newblock Why asymmetric molecular coupling to electrodes cannot be at work in
  real molecular rectifiers.
\newblock {\em Phys. Rev. B} {\bf 2021}, {\em 103},~195408.
\newblock
  {\changeurlcolor{black}\href{https://doi.org/https://dx.doi.org/10.1103/PhysRevB.103.195408}{\detokenize{https://dx.doi.org/10.1103/PhysRevB.103.195408}}}.

\bibitem[Sommerfeld and Bethe(1933)]{Sommerfeld:33}
Sommerfeld, A.; Bethe, H.
\newblock Elektronentheorie der Metalle. In {\em Handbuch der Physik}; Scheel, G., Ed.; Julius-Springer: Berlin,  Germany, 1933; Volume 24, p. 446.

\bibitem[Desjonqueres and Spanjaard(1996)]{desjonqueres:96}
Desjonqueres, M.C.; Spanjaard, D.
\newblock {\em Concepts in Surface Physics}, 2nd ed.; Springer: Berlin/Heidelberg, Germany; New York, NY, USA, 1996.

\bibitem[Neaton \em{et~al.}(2006)Neaton, Hybertsen, and Louie]{Neaton:06}
Neaton, J.B.; Hybertsen, M.S.; Louie, S.G.
\newblock Renormalization of Molecular Electronic Levels at Metal-Molecule
  Interfaces.
\newblock {\em Phys. Rev. Lett.} {\bf 2006}, {\em 97},~216405.
\newblock
  {\changeurlcolor{black}\href{https://doi.org/10.1103/PhysRevLett.97.216405}{{https://doi.org/10.1103/PhysRevLett.97.216405}}}.

\bibitem[B\^aldea(2014{\natexlab{a}})]{Baldea:2014a}
B\^aldea, I.
\newblock Single-Molecule Junctions Based on Bipyridine: Impact of an Unusual
  Reorganization on the Charge Transport.
\newblock {\em J. Phys. Chem. C} {\bf 2014}, {\em 118},~8676--8684.
  \href{http://xxx.lanl.gov/abs/http://pubs.acs.org/doi/pdf/10.1021/jp412675k}
  {\changeurlcolor{black}\href{https://doi.org/10.1021/jp412675k}{{https://doi.org/10.1021/jp412675k}}}.

\bibitem[B\^aldea(2014{\natexlab{b}})]{Baldea:2014e}
B\^aldea, I.
\newblock Quantifying the Relative Molecular Orbital Alignment for Molecular
  Junctions with Similar Chemical Linkage to Electrodes.
\newblock {\em Nanotechnology} {\bf 2014}, {\em 25},~455202.
\newblock
  {\changeurlcolor{black}\href{https://doi.org/10.1088/0957-4484/25/45/455202}{\detokenize{https://doi.org/10.1088/0957-4484/25/45/455202}}}.

\bibitem[Abramowitz and Stegun(1964)]{AbramowitzStegun:64}
Abramowitz, M.; Stegun, I.A. (Eds.)
\newblock {\em Handbook of Mathematical Functions with Formulas, Graphs, and
  Mathematical Tables}; National Bureau of Standards Applied Mathematics
  Series; U.S. Government Printing Office: Washington, DC,  USA, 1964.

\bibitem[B\^aldea(2012)]{Baldea:2012g}
B\^aldea, I.
\newblock Interpretation of Stochastic Events in Single-Molecule Measurements
  of Conductance and Transition Voltage Spectroscopy.
\newblock {\em J. Am. Chem. Soc.} {\bf 2012}, {\em 134},~7958--7962.
  \href{http://xxx.lanl.gov/abs/http://pubs.acs.org/doi/pdf/10.1021/ja302248h}
\newblock
  {\changeurlcolor{black}\href{https://doi.org/10.1021/ja302248h}{{https://doi.org/10.1021/ja302248h}}}.

\bibitem[Sedghi \em{et~al.}(2011)Sedghi, Garcia-Suarez, Esdaile, Anderson,
  Lambert, Martin, Bethell, Higgins, Elliott, Bennett, Macdonald, and
  Nichols]{Lambert:11}
Sedghi, G.; Garcia-Suarez, V.M.; Esdaile, L.J.; Anderson, H.L.; Lambert, C.J.;
  Martin, S.; Bethell, D.; Higgins, S.J.; Elliott, M.; Bennett, N.;  et~al.
\newblock Long-Range Electron Tunnelling in Oligo-Porphyrin Molecular Wires.
\newblock {\em Nat. Nanotechnol.} {\bf 2011}, {\em 6},~517--523.
\newblock
  {\changeurlcolor{black}\href{https://doi.org/10.1038/nnano.2011.111}{{https://doi.org/10.1038/nnano.2011.111}}}.

\bibitem[Smith \em{et~al.}(2018)Smith, Xie, B\^aldea, and
  Frisbie]{Baldea:2018a}
Smith, C.E.; Xie, Z.; B\^aldea, I.; Frisbie, C.D.
\newblock Work Function and Temperature Dependence of Electron Tunneling
  through an N-Type Perylene Diimide Molecular Junction with Isocyanide Surface
  Linkers.
\newblock {\em Nanoscale} {\bf 2018}, {\em 10},~964--975.
\newblock
  {\changeurlcolor{black}\href{https://doi.org/10.1039/C7NR06461F}{{https://doi.org/10.1039/C7NR06461F}}}.

\bibitem[Jahnke and Emde(1945)]{JahnkeEmde:45}
Jahnke, E.; Emde, F.
\newblock {\em Tables of Functions with Formulae and Curves}, 4th ed.; For the Riemann zeta function; Dover Publications: {New York, NY, USA}, 1945; p. 269.


\bibitem[Ashcroft and Mermin(1976)]{AshcroftMermin}
Ashcroft, N.W.; Mermin, N.D.
\newblock {\em Solid State Physics}; Saunders College Publishing: New York, NY, USA,
  1976; pp. 20--23, 52.

\bibitem[Poot \em{et~al.}(2006)Poot, Osorio, O'Neill, Thijssen, Vanmaekelbergh,
  van Walree, Jenneskens, and van~der Zant]{Poot:06}
Poot, M.; Osorio, E.; O'Neill, K.; Thijssen, J.M.; Vanmaekelbergh, D.; van
  Walree, C.A.; Jenneskens, L.W.; van~der Zant, H.S.J.
\newblock Temperature Dependence of Three-Terminal Molecular Junctions with
  Sulfur End-Functionalized Tercyclohexylidenes.
\newblock {\em Nano Lett.} {\bf 2006}, {\em 6},~1031--1035.
  \href{http://xxx.lanl.gov/abs/http://pubs.acs.org/doi/pdf/10.1021/nl0604513}
\newblock
  {\changeurlcolor{black}\href{https://doi.org/10.1021/nl0604513}{{https://doi.org/10.1021/nl0604513}}}.

\bibitem[Heimbuch \em{et~al.}(2012)Heimbuch, Wu, Kumar, Poelsema, Sch\"on,
  Vancso, and Zandvliet]{Zandvliet:12}
Heimbuch, R.; Wu, H.; Kumar, A.; Poelsema, B.; Sch\"on, P.; Vancso, G.J.;
  Zandvliet, H.J.W.
\newblock Variable-Temperature Study of the Transport Through a Single
  Octanethiol Molecule.
\newblock {\em Phys. Rev. B} {\bf 2012}, {\em 86},~075456.
\newblock
  {\changeurlcolor{black}\href{https://doi.org/10.1103/PhysRevB.86.075456}{{https://doi.org/10.1103/PhysRevB.86.075456}}}.

\bibitem[Asadi \em{et~al.}(2013)Asadi, Kronemeijer, Cramer, Jan Anton~Koster,
  Blom, and de~Leeuw]{Asadi:13}
Asadi, K.; Kronemeijer, A.J.; Cramer, T.; Jan Anton~Koster, L.; Blom, P.W.M.;
  de~Leeuw, D.M.
\newblock Polaron hopping mediated by nuclear tunnelling in semiconducting
  polymers at high carrier density.
\newblock {\em Nat. Commun.} {\bf 2013}, {\em 4},~1710.
\newblock
  {\changeurlcolor{black}\href{https://doi.org/10.1038/ncomms2708}{{https://doi.org/10.1038/ncomms2708}}}.

\bibitem[Xiang \em{et~al.}(2016)Xiang, Hines, Palma, Lu, Mujica, Ratner, Zhou,
  and Tao]{Tao:16b}
Xiang, L.; Hines, T.; Palma, J.L.; Lu, X.; Mujica, V.; Ratner, M.A.; Zhou, G.;
  Tao, N.
\newblock Non-Exponential Length Dependence of Conductance in Iodide-Terminated
  Oligothiophene Single-Molecule Tunneling Junctions.
\newblock {\em J. Am. Chem. Soc.} {\bf 2016}, {\em 138},~679--687.
  \href{http://xxx.lanl.gov/abs/http://dx.doi.org/10.1021/jacs.5b11605}
  {\changeurlcolor{black}\href{https://doi.org/10.1021/jacs.5b11605}{{https://doi.org/10.1021/jacs.5b11605}}}.

\bibitem[McCreery(2016)]{McCreery:16a}
McCreery, R.L.
\newblock Effects of Electronic Coupling and Electrostatic Potential on Charge
  Transport in Carbon-Based Molecular Electronic Junctions.
\newblock {\em Beilstein J. Nanotechnol.} {\bf 2016}, {\em 7},~32--46.
\newblock
  {\changeurlcolor{black}\href{https://doi.org/10.3762/bjnano.7.4}{{https://doi.org/10.3762/bjnano.7.4}}}.

\bibitem[Kumar \em{et~al.}(2016)Kumar, Pasula, Lim, and Nijhuis]{Nijhuis:16d}
Kumar, K.S.; Pasula, R.R.; Lim, S.; Nijhuis, C.A.
\newblock Long-Range Tunneling Processes across Ferritin-Based Junctions.
\newblock {\em Adv. Mater.} {\bf 2016}, {\em 28},~1824--1830.
  \href{http://xxx.lanl.gov/abs/https://onlinelibrary.wiley.com/doi/pdf/10.1002/adma.201504402}
  {\changeurlcolor{black}\href{https://doi.org/https://doi.org/10.1002/adma.201504402}{\detokenize{https://doi.org/10.1002/adma.201504402}}}.

\bibitem[Xin \em{et~al.}(2017)Xin, Jia, Wang, Wang, Li, Gong, Zhang, Zhu, and
  Guo]{Guo:17}
Xin, N.; Jia, C.; Wang, J.; Wang, S.; Li, M.; Gong, Y.; Zhang, G.; Zhu, D.;
  Guo, X.
\newblock Thermally Activated Tunneling Transition in a Photoswitchable
  Single-Molecule Electrical Junction.
\newblock {\em J. Phys. Chem. Lett.} {\bf 2017}, {\em 8},~2849--2854.
  \href{http://xxx.lanl.gov/abs/https://doi.org/10.1021/acs.jpclett.7b01063}
  {\changeurlcolor{black}\href{https://doi.org/10.1021/acs.jpclett.7b01063}{{https://doi.org/10.1021/acs.jpclett.7b01063}}}.

\bibitem[Morteza~Najarian and McCreery(2017)]{McCreery:17a}
Morteza~Najarian, A.; McCreery, R.L.
\newblock Structure Controlled Long-Range Sequential Tunneling in Carbon-Based
  Molecular Junctions.
\newblock {\em ACS Nano} {\bf 2017}, {\em 11},~3542--3552.
  \href{http://xxx.lanl.gov/abs/http://dx.doi.org/10.1021/acsnano.7b00597}
  {\changeurlcolor{black}\href{https://doi.org/10.1021/acsnano.7b00597}{{https://doi.org/10.1021/acsnano.7b00597}}}.

\bibitem[Xin \em{et~al.}(2021)Xin, Hu, Al~Sabea, Zhang, Zhou, Meng, Jia, Gong,
  Li, Ke, He, Selvanathan, Norel, Ratner, Liu, Xiao, Rigaut, Guo, and
  Guo]{Guo:21}
Xin, N.; Hu, C.; Al~Sabea, H.; Zhang, M.; Zhou, C.; Meng, L.; Jia, C.; Gong,
  Y.; Li, Y.; Ke, G.;  et~al.
\newblock Tunable Symmetry-Breaking-Induced Dual Functions in Stable and
  Photoswitched Single-Molecule Junctions.
\newblock {\em J. Am. Chem. Soc.} {\bf 2021}, {\em 143},~20811--20817.
  \href{http://xxx.lanl.gov/abs/https://doi.org/10.1021/jacs.1c08997}
  {\changeurlcolor{black}\href{https://doi.org/10.1021/jacs.1c08997}{{https://doi.org/10.1021/jacs.1c08997}}}.

\bibitem[Haaland and Nilsson(1968)]{Haaland:68}
Haaland, A.; Nilsson, J.E.
\newblock The Determination of Barriers to Internal Rotation by Means of
  Electron Diffraction. Ferrocene and Ruthenocene.
\newblock {\em Acta Chem. Scand.} {\bf 1968}, {\em 22},~2653--2670.
\newblock
  {\changeurlcolor{black}\href{https://doi.org/10.3891/acta.chem.scand.22-2653}{{https://doi.org/10.3891/acta.chem.scand.22-2653}}}.

\bibitem[Coriani \em{et~al.}(2006)Coriani, Haaland, Helgaker, and
  J{o}rgensen]{Coriani:06}
Coriani, S.; Haaland, A.; Helgaker, T.; J{o}rgensen, P.
\newblock The Equilibrium Structure of Ferrocene.
\newblock {\em ChemPhysChem} {\bf 2006}, {\em 7},~245--249.
  \href{http://xxx.lanl.gov/abs/https://chemistry-europe.onlinelibrary.wiley.com/doi/pdf/10.1002/cphc.200500339}
  {\changeurlcolor{black}\href{https://doi.org/https://doi.org/10.1002/cphc.200500339}{\detokenize{https://doi.org/10.1002/cphc.200500339}}}.

\bibitem[B\^aldea(2013)]{Baldea:2013b}
B\^aldea, I.
\newblock Transition Voltage Spectroscopy Reveals Significant Solvent Effects
  on Molecular Transport and Settles an Important Issue in Bipyridine-Based
  Junctions.
\newblock {\em Nanoscale} {\bf 2013}, {\em 5},~9222--9230.
\newblock
  {\changeurlcolor{black}\href{https://doi.org/10.1039/C3NR51290H}{{https://doi.org/10.1039/C3NR51290H}}}.

\bibitem[Xie \em{et~al.}(2015)Xie, B\^aldea, Smith, Wu, and
  Frisbie]{Baldea:2015d}
Xie, Z.; B\^aldea, I.; Smith, C.; Wu, Y.; Frisbie, C.D.
\newblock Experimental and Theoretical Analysis of Nanotransport in
  Oligophenylene Dithiol Junctions as a Function of Molecular Length and
  Contact Work Function.
\newblock {\em ACS Nano} {\bf 2015}, {\em 9},~8022--8036.
  \href{http://xxx.lanl.gov/abs/http://dx.doi.org/10.1021/acsnano.5b01629}
  {\changeurlcolor{black}\href{https://doi.org/10.1021/acsnano.5b01629}{{https://doi.org/10.1021/acsnano.5b01629}}}.

\bibitem[Xie \em{et~al.}(2019{\natexlab{a}})Xie, B\^aldea, and
  Frisbie]{Baldea:2019d}
Xie, Z.; B\^aldea, I.; Frisbie, C.D.
\newblock{ Determination of Energy Level Alignment in Molecular Tunnel Junctions
  by Transport and Spectroscopy: Self-Consistency for the Case of
  Oligophenylene Thiols and Dithiols on Ag, Au, and Pt Electrodes.}

\newblock {\em J. Am. Chem. Soc.} {\bf 2019}, {\em 141},~3670--3681.
  \href{http://xxx.lanl.gov/abs/https://doi.org/10.1021/jacs.8b13370}
\newblock
  {\changeurlcolor{black}\href{https://doi.org/10.1021/jacs.8b13370}{{https://doi.org/10.1021/jacs.8b13370}}}.

\bibitem[Xie \em{et~al.}(2019{\natexlab{b}})Xie, B\^aldea, and
  Frisbie]{Baldea:2019h}
Xie, Z.; B\^aldea, I.; Frisbie, C.D.
\newblock {Energy Level Alignment in Molecular Tunnel Junctions by Transport and
  Spectroscopy: Self-Consistency for the Case of Alkyl Thiols and Dithiols on
  Ag, Au, and Pt Electrodes.}
\newblock {\em J. Am. Chem. Soc.} {\bf 2019}, {\em 141},~18182--18192.
  \href{http://xxx.lanl.gov/abs/https://doi.org/10.1021/jacs.9b08905}
  {\changeurlcolor{black}\href{https://doi.org/10.1021/jacs.9b08905}{{https://doi.org/10.1021/jacs.9b08905}}}.

\bibitem[B\^aldea and K\"oppel(2012)]{Baldea:2012c}
B\^aldea, I.; K\"oppel, H.
\newblock Evidence on single-molecule transport in electrostatically-gated
  molecular transistors.
\newblock {\em Phys. Lett. A} {\bf 2012}, {\em 376},~1472--1476.
\newblock
  {\changeurlcolor{black}\href{https://doi.org/10.1016/j.physleta.2012.03.021}{{https://doi.org/10.1016/j.physleta.2012.03.021}}}.

\bibitem[del Barco()]{delBarco}
del Barco, E.
\newblock {Private commuication}

\bibitem[Demissie \em{et~al.}(2016)Demissie, Haugstad, and
  Frisbie]{Frisbie:16e}
Demissie, A.T.; Haugstad, G.; Frisbie, C.D.
\newblock Quantitative Surface Coverage Measurements of Self-Assembled
  Monolayers by Nuclear Reaction Analysis of Carbon-12.
\newblock {\em J. Phys. Chem. Lett.} {\bf 2016}, {\em 7},~3477--3481.
  \href{http://xxx.lanl.gov/abs/http://dx.doi.org/10.1021/acs.jpclett.6b01363}
  {\changeurlcolor{black}\href{https://doi.org/10.1021/acs.jpclett.6b01363}{{https://doi.org/10.1021/acs.jpclett.6b01363}}}.

\bibitem[Xie \em{et~al.}(2017)Xie, B\^aldea, Demissie, Smith, Wu, Haugstad, and
  Frisbie]{Baldea:2017e}
Xie, Z.; B\^aldea, I.; Demissie, A.T.; Smith, C.E.; Wu, Y.; Haugstad, G.;
  Frisbie, C.D.
\newblock Exceptionally Small Statistical Variations in the Transport
  Properties of Metal-Molecule-Metal Junctions Composed of 80 Oligophenylene
  Dithiol Molecules.
\newblock {\em J. Am. Chem. Soc.} {\bf 2017}, {\em 139},~5696--5699.
  \href{http://xxx.lanl.gov/abs/https://dx.doi.org/10.1021/jacs.7b01918}
  {\changeurlcolor{black}\href{https://doi.org/10.1021/jacs.7b01918}{{https://doi.org/10.1021/jacs.7b01918}}}.

\bibitem[Li()]{YuanLi}
Li, Y.
\newblock
{Private commuication}

\bibitem[Choi \em{et~al.}(2008)Choi, Kim, and Frisbie]{Choi:08}
Choi, S.H.; Kim, B.; Frisbie, C.D.
\newblock Electrical Resistance of Long Conjugated Molecular Wires.
\newblock {\em Science} {\bf 2008}, {\em 320},~1482--1486.
  \href{http://xxx.lanl.gov/abs/http://www.sciencemag.org/cgi/reprint/320/5882/1482.pdf}
  {\changeurlcolor{black}\href{https://doi.org/10.1126/science.1156538}{{https://doi.org/10.1126/science.1156538}}}.

\bibitem[Hines \em{et~al.}(2010)Hines, Diez-Perez, Hihath, Liu, Wang, Zhao,
  Zhou, M\"ullen, and Tao]{Tao:10}
Hines, T.; Diez-Perez, I.; Hihath, J.; Liu, H.; Wang, Z.S.; Zhao, J.; Zhou, G.;
  M\"ullen, K.; Tao, N.
\newblock Transition from Tunneling to Hopping in Single Molecular Junctions by
  Measuring Length and Temperature Dependence.
\newblock {\em J. Am. Chem. Soc.} {\bf 2010}, {\em 132},~11658--11664.
  \href{http://xxx.lanl.gov/abs/http://pubs.acs.org/doi/pdf/10.1021/ja1040946}
  {\changeurlcolor{black}\href{https://doi.org/10.1021/ja1040946}{{https://doi.org/10.1021/ja1040946}}}.

\bibitem[Shklovskii and Efros(1984)]{Shklovskii:84}
Shklovskii, B.I.; Efros, A.L. Variable-Range Hopping Conduction.
\newblock In {\em Electronic Properties of Doped Semiconductors}; Springer: Berlin/Heidelberg, Germany, 1984; pp. 202--227.
\newblock
  {\changeurlcolor{black}\href{https://doi.org/10.1007/978-3-662-02403-4_9}{{https://doi.org/10.1007/978-3-662-02403-4\_9}}}.

\bibitem[Maugis(1992)]{Maugis:92}
Maugis, D.
\newblock Adhesion of Spheres: The JKR-DMT Transition Using a Dugdale Model.
\newblock {\em J. Colloid. Interf. Sci.} {\bf 1992}, {\em 150},~243--269.
\newblock
  {\changeurlcolor{black}\href{https://doi.org/https://dx.doi.org/10.1016/0021-9797(92)90285-T}{\detokenize{https://dx.doi.org/10.1016/0021-9797(92)90285-T}}}.

\bibitem[Johnson(1985)]{Johnson:85}
Johnson, K.L.
\newblock {\em Contact Mechanics}; Cambridge University Press: {Cambridge, UK,} 
 1985.
\newblock
  {\changeurlcolor{black}\href{https://doi.org/10.1017/CBO9781139171731}{{https://doi.org/10.1017/CBO9781139171731}}}.

\bibitem[Haugstad(2012)]{Haugstad:12}
Haugstad, G.
\newblock {\em Atomic Force Microscopy}; John Wiley \& Sons:  {Hoboken, NJ, USA,} 
2012.
  \href{http://xxx.lanl.gov/abs/https://onlinelibrary.wiley.com/doi/pdf/10.1002/9781118360668}
  {\changeurlcolor{black}\href{https://doi.org/10.1002/9781118360668}{{https://doi.org/10.1002/9781118360668}}}.

\bibitem[B\^aldea(2021)]{Baldea:2021a}
B\^aldea, I.
\newblock Self-assembled monolayers of oligophenylenes stiffer than steel and
  silicon, possibly even stiffer than Si3N4.
\newblock {\em Appl. Surf. Sci. Adv.} {\bf 2021}, {\em 5},~100094.
\newblock
  {\changeurlcolor{black}\href{https://doi.org/https://doi.org/10.1016/j.apsadv.2021.100094}{\detokenize{https://doi.org/10.1016/j.apsadv.2021.100094}}}.

\bibitem[B\^aldea(2013)]{Baldea:2013d}
B\^aldea, I.
\newblock Important Insight into Electron Transfer in Single-Molecule Junctions
  Based on Redox Metalloproteins from Transition Voltage Spectroscopy.
\newblock {\em J. Phys. Chem. C} {\bf 2013}, {\em 117},~25798--25804.
  \href{http://xxx.lanl.gov/abs/http://pubs.acs.org/doi/pdf/10.1021/jp408873c}
  {\changeurlcolor{black}\href{https://doi.org/10.1021/jp408873c}{{https://doi.org/10.1021/jp408873c}}}.

\end{thebibliography}
\end{document}